\documentclass[11pt,a4paper]{article}
\pdfoutput=1
\usepackage{jheppub}
\usepackage{amsmath}
\usepackage{amssymb}
\usepackage{amsfonts}
\usepackage{bbm} 
\usepackage{graphics}  
\usepackage{graphicx} 
\usepackage{subfigure}
\usepackage{hhline}
\usepackage{longtable}  
\usepackage{multirow} 
\usepackage{dsfont}  
\usepackage[all,cmtip]{xy} 

\newcommand{\cO}{\mathcal{O}}

\newcommand{\cE}{\mathcal{E}}

\newcommand{\cN}{\mathcal{N}}

\newcommand{\cF}{\mathcal{F}}

\newcommand{\cV}{\mathcal{V}}
\newcommand{\cM}{\mathcal{M}}
\newcommand{\cQ}{\mathcal{Q}}
\newcommand{\cZ}{\mathcal{Z}}

\newcommand{\bfa}{{\bf a}}
\newcommand{\bsa}{{\boldsymbol a}}

\newcommand{\bc}{{\bf c}}
\newcommand{\bC}{{\bf C}}
\newcommand{\bfd}{{\bf d}}
\newcommand{\bd}{{\boldsymbol d}}
\newcommand{\bff}{{\bf f}}
\newcommand{\bfe}{{\bf e}}
\newcommand{\bg}{{\boldsymbol g}}
\newcommand{\bfk}{\mathbf{k}}

\def\IZ{{\mathbb Z}}
\def\IR{{\mathbb R}}

\def\IP{{\mathbb P}}

\def\IF{{\mathbb F}}
\def\IB{{\mathbb B}}
\newcommand{\bP}{\mathbb{P}}

\newcommand{\sgn}{{\rm sgn}}

\newcommand{\half}{\frac12}

\newcommand{\be}{\begin{equation}}
\newcommand{\ee}{\end{equation}}
\newcommand{\ba}{\begin{aligned}}
\newcommand{\ea}{\end{aligned}}

\newcommand{\ben}{\begin{eqnarray}}
\newcommand{\een}{\end{eqnarray}}

\newcommand{\noi}{\noindent}
\newcommand{\im}{\mathrm{Im}}
\newcommand{\Tr}{\mathrm{Tr}}
\let\non\nonumber

\title{A 5d/2d/4d correspondence}
\author[a]{Babak Haghighat,}
\author[b,c]{Jan Manschot,}
\author[d]{Stefan Vandoren}
\affiliation[a]{Jefferson Physical Laboratory, Harvard University, Cambridge, MA 02138, USA}
\affiliation[b]{Bethe Center for Theoretical Physics, Physikalisches Institut, Universit\"at Bonn, Nussallee 12, 53115 Bonn, Germany}
\affiliation[c]{Max-Planck-Institut f\"ur Mathematik, Vivatsgasse 7, 53111 Bonn, Germany}
\affiliation[d]{Institute for Theoretical Physics and Spinoza Institute, Utrecht University, 3508 TD Utrecht, The Netherlands}
\emailAdd{babak@physics.harvard.edu, manschot@uni-bonn.de, S.J.G.Vandoren@uu.nl}

\abstract
{

We propose a correspondence between two-dimensional $(0,4)$
sigma models with target space the moduli spaces of $r$ monopoles, and
four-dimensional $\cN=4$, $U(r)$ Yang-Mills theory on del Pezzo surfaces.
In particular, the two- and four-dimensional BPS partition functions are
  argued to be equal. The correspondence relies on insights from
  five-dimensional supersymmetric gauge theory and its geometric engineering in M-theory, hence the name
  ``5d/2d/4d  correspondence''. We provide various tests of our
  proposal. The most stringent ones are for $r=1$, for which we prove
  the equality of partition functions.

}

\begin{document}
\hfill BONN-TH-2012-25

\hfill ITP-UU-12/35

\hfill SPIN-12/33

\maketitle


\section{Introduction and summary}

The main aim of this paper is to present a correspondence between two
different theories, namely a $(0,4)$ non-linear sigma model with as target space the moduli space of magnetic monopoles 
and $\cN=4$ super-Yang-Mills theory on a del Pezzo surface. This correspondence, anticipated in \cite{arXiv:1107.2847}, follows from combining various observations and studies in the existing literature. The close relation between
two-dimensional conformal field theory and $\cN=4$ super Yang-Mills was
already noticed in the seminal paper \cite{Vafa:1994tf}. The relation
was put on a much firmer footing after the discovery of the duality
between five-branes wrapped on $K3$ and the fundamental heterotic string
\cite{Witten:1995ex, Harvey:1995rn, Sen:1995cj}, and similarly
${1\over 2}K3$ and E-strings \cite{Minahan:1998vr}. A systematic reduction of  the 
degrees of freedom of M5-branes on more general four-manifolds
embedded in a {\it compact} Calabi-Yau three-fold to a $(0,4)$ CFT was initiated by
Maldacena, Strominger and Witten \cite{Maldacena:1997de}. We will refer to this as the MSW CFT. 

In this paper we consider the class of {\it non-compact} Calabi-Yau
three-folds $X$ which are locally canonical bundles over del Pezzo surfaces $P$. In particular, we consider M-theory compactified on $X\times T^2$, with M5-branes wrapped on $P\times T^2$. This leads to the 
mentioned correspondence between monopole moduli spaces and Yang-Mills
theory, since the degrees of freedom of $r$ coincident M5-branes
reduced to four dimensions equal $U(r)$, $\cN=4$ super Yang-Mills on $P$, and on
the other hand geometric engineering of quantum field theories in string theory \cite{Morrison:1996xf,Douglas:1996xp, Intriligator:1997pq, Katz:1996fh} relate the M5-branes on $P$ to magnetic strings. The worldvolume theory of the magnetic string is a sigma model that
appears to be of a rather different form compared to the MSW CFT
\cite{Maldacena:1997de, Minasian:1999qn}.

Due to the important role of M5-branes for the correspondence, our introduction and motivation will 
 continue with a short discussion on M5-branes and their
partition functions, after which we propose the correspondence that
leads to the main conjecture in formula \eqref{correspondence}. At the
end, in Section \eqref{5d2d4d}, we give a first explanation of the
conjecture using 5d $\cN=1$ supersymmetric gauge theories.

\subsection{Motivation: M5-branes}

Understanding the M5-brane worldvolume theory and formulating a
consistent action for this theory has been a long standing open
problem. First steps towards the solution were taken in
\cite{Witten:1996hc}. One of the main difficulties in the description
is the existence of a self-dual three-form $H$, which is the field
strength of the five-brane two-form $B$, and for which no Lagrangian
formulation is available. This fact also makes it difficult to define
a convenient partition function for the M5-brane theory. However, in
\cite{Witten:1996hc} it was noticed that the partition function is a
certain section of a line bundle over the intermediate Jacobian  
$J_W = H^3(W,\IR)/H^3(W,\IZ)$ where $W$ is the six-manifold on which
the five-brane is wrapped. This can be traced back to the coupling of
the two-form $B$ to the M-theory three-form $C$ and can be understood
in an intuitive manner as follows. Consider the specific case of $W =
\Sigma \times \IP^2$, with $\Sigma$ a Riemann surface whose
characteristic length is much larger than that of $\IP^2$. Then the theory of the chiral
two-form reduces to a chiral scalar on $\Sigma$. This chiral scalar
can now be coupled to a $U(1)$ gauge field $A$ which when setting the
curvature $F = dA$ to zero defines modulo gauge transformations a
point on $H^1(\Sigma,\IR)/H^1(\Sigma,\IZ)$. In order for the gauge
field to couple to the chiral part of the scalar only, the Lagrangian
contains a term which breaks gauge invariance. Thus the partition
function $\cZ$, defined by taking the path integral over the scalar,
is not a function of $A$ but rather a section of a line bundle over
$H^1(\Sigma,\IR)/H^1(\Sigma,\IZ)$. This point of view is convenient
when one wants to establish contact with wave functions and a
background independent interpretation of partition functions
\cite{Witten:1993ed}. It also naturally leads to a description of the
partition function in terms of theta-functions which can be
interpreted as sections of line bundles on the Jacobian.  

A more general situation where one considers instead of $\IP^2$ an
arbitrary K\"ahler manifold $P$ and takes $\Sigma$ to be the torus
$T^2$ has been extensively analyzed in \cite{Maldacena:1997de,Denef:2007vg,
  Gaiotto:2006wm, Minasian:1999qn, deBoer:2006vg}. In this case the
reduction of the two-form will give rise to left-moving and
right-moving chiral scalars whose numbers are determined by the
self-dual and anti-self-dual harmonic two-forms on $P$. One then
considers a situation where $P$ is embedded into a Calabi-Yau
threefold $X$ and takes the embedding to be holomorphic in order to
preserve supersymmetry. The deformation degrees of freedom of the
five-brane together with the chiral scalars from the reduction of the
two-form form a $(0,4)$ CFT whose partition function is a modular form
with its modular parameter being the complex structure $\tau$ of
$T^2$. In \cite{Minasian:1999qn} it is argued that there is an
underlying sigma model for this CFT whose target space $\cE$ is a bundle of
the form $\cV \rightarrow {\cM}$ where $\cV$ is a
vector-bundle with real rank $b_2^-(P) - b_2^+(P) - (h^{1,1}(X)-2)$
and the base-manifold ${\cM}$ carries a hyperk\"ahler
structure. One drawback of present day constructions of the bundle
$\cE$ is that they are only known for the case of a single
M5-brane. Thus a major question to answer is what happens if one
wraps an arbitrary number of five-branes around $P$. We will provide an answer to this question
when $P$ is a del Pezzo surface and $X$ is non-compact.

Yet another viewpoint is obtained by turning the situation
around. Instead of taking the size of $P$ to be small one considers a
limit where the characteristic length of $P$ is much larger than that of $T^2$. Reduction of the degrees
of freedom of $r$ coincident five-branes to $P$ leads to topologically twisted
$\cN=4$ supersymmetric $U(r)$ Yang-Mills theory 
\cite{Vafa:1994tf, Minahan:1998vr}. The key point here is that the complex
structure $\tau$ of $T^2$, will take over the role of the
complexified gauge coupling of the Yang-Mills theory. The $U(r)$
partition functions can be evaluated in various cases. One example is
the case where $P$ is an elliptic fibration, such that one can invoke twice
T-duality along the elliptic fiber to map the Yang-Mills partition
function for arbitrary rank $r$ to topological string free energies \cite{Minahan:1998vr,
  Alim:2010cf, Klemm:2012sx}. More generally,
Yang-Mills partition functions for arbitrary rank $r$ can be evaluated
for all rational surfaces using algebraic-geometric techniques as wall-crossing and
blow-up formulas \cite{Gottsche:1990, Yoshioka:1994, Yoshioka:1995, Manschot:2011dj, Manschot:2011ym,
  Manschot:2010nc}. 

\subsection{Statement of the conjecture}

As mentioned above, the $(0,4)$ CFT on $T^2$
and the $\cN=4$ SYM on $P$ are related by the effective action of the
M5-brane world-volume theory. Indeed, at the level of partition
functions one can consider the elliptic genus of the two-dimensional
theory and a particular topologically twisted version of Yang-Mills
theory\footnote{As will be discussed in Section 3, the reduction of
  the five-brane theory along $T^2$ will automatically lead to a
  topological theory on $P$.}. Both quantities will be index-like and
therefore their dependence on the volumes of $T^2$ and $P$ enters only
in a trivial way. Moreover, both quantities are modular functions,
with equal modular weight $(0,2)$. This suggests that the underlying six-dimensional
theory will ultimately connect the two and one expects a relation of
the following form 
\begin{equation} \label{correspondence}
	\cZ^{(r)}_{(0,4)~\textrm{CFT}} = \cZ^{(r)}_{\textrm{SYM}}\ ,
\end{equation}
where $r$ denotes the number of M5-branes wrapped around $T^2 \times
P$. The correspondence appeared in this explicit form for the first
time in \cite{Minahan:1998vr}. It is also reminiscent to the 2d/4d dualities studied in
\cite{Alday:2009aq}. But note that in the present case both 2d and 4d theories
are supersymmetric whereas in \cite{Alday:2009aq} the dual 2d theory is nonsupersymmetric
Liouville theory. 

The task of this paper is, following up on \cite{arXiv:1107.2847}, to shed new light on the
correspondence (\ref{correspondence}) and in particular clarify the
nature of the underlying sigma model which gives rise to the $(0,4)$
CFT. To do this, we restrict to the case where $P$ is a del Pezzo
surface and state a conjecture for (\ref{correspondence}) for
arbitrary high $r$. 

Del Pezzo surfaces are K\"ahler manifolds of complex
dimension two with positive anti-canonical class, which makes them
rigid inside the Calabi-Yau. Such surfaces are either $\IF_0 = \IP^1 \times \IP^1$, $\IF_1$ or
blow-ups of these. Note that blowing-up a point of $\IF_0$ or $\IF_1$
give topologically equal manifolds. We define $\IB_0 := \IF_0$ and $\IB_n$ as the blow-up of $\IB_0$ at
$n$ points. We will not consider in detail the case $P = \IF_1$, since
the all qualitative aspects of the correspondence can be explained
using the class $\IB_n$. For these surfaces $b_2^+(P) = 1$, such
that interesting wall-crossing effects arise in the $\cN=4$ Yang-Mills
theory on $P$. 

Let us elaborate further on the formulation of our conjecture. It says that wrapping
$r$ M5-branes around $T^2 \times \IB_n$ and taking the size of $T^2$
to be much larger than that of $\IB_n$ will give rise to a $(0,4)$
sigma model with target space being the moduli space of $r$ $SU(2)$
magnetic monopoles in the presence of $2 r n$ fermionic zero
modes coming from adding massless flavor fermions in the fundamental representation 
to the $SU(2)$ gauge theory. Hence the number of blow-ups is identified with the number of 
flavors, $N_f=n$. The structure of the target space of the $(0,4)$ sigma model has the following bundle structure \cite{Manton:1993aa}:
\begin{equation}
\begin{array}{cc}
	O(r) \times SO(2N_f) & ~\\
	\downarrow & ~ \\
	\cM_r & = \quad \IR^3 \times \frac{S^1 \times \widetilde{\cM}_r}{\IZ_r}~.
\end{array}
\end{equation}
In the above, $\cM_r$ denotes the moduli space of magnetic monopoles
of charge $r$ in pure $SU(2)$ Yang-Mills theory. The factor $\IR^3$
can be understood as the zero modes corresponding to the center of
mass motion in space, $S^1$ represents the unbroken $U(1)$ charge of
the monopoles and $\widetilde{\cM}_r$ captures the relative moduli
space. Over the base $\cM_r$, there is a $O(r)$ vector bundle with transition functions in the orthogonal group $O(r)$.
This bundle is also called the Index bundle associated to the magnetic monopole moduli space, as discussed in  
\cite{Manton:1993aa}. Finally, there is an additional isometric action of the flavor group $SO(2N_f)$ which we discuss in the following sections below.

We now claim that the elliptic genus, as defined in
Section 4, of the sigma model corresponding to this target space is
equal to the partition function of topologically twisted $U(r)$ $\cN=4$
SYM on $\IB_{n=N_f}$. The topological twist is the one introduced in
\cite{Vafa:1994tf} and as discussed there the path-integral of the
twisted Yang-Mills theory localizes on instanton configurations.   

\subsection{Explanation of the conjecture: a 5d/2d/4d correspondence}\label{5d2d4d}

We turn to an explanation of the main idea behind the claimed
correspondence. Consider five-dimensional $\cN=1$ supersymmetric gauge
theory with $SU(2)$ gauge group whose properties have been first
discussed in \cite{Seiberg:1996bd}. This theory can be geometrically
engineered, as was done in \cite{Morrison:1996xf, Intriligator:1997pq,Douglas:1996xp}
(see also \cite{arXiv:1107.2847}), by compactifying M-theory on a
Calabi-Yau threefold which is locally the canonical bundle over the
del Pezzo surface $\IB_{N_f}$. Here, $N_f$ denotes the number of
flavors in the fundamental of $SU(2)$. These are accompanied with a
$U(1)^{N_f}$ flavor-symmetry which gets enhanced to $SO(2N_f)$ in the
massless case. The spectrum of the five-dimensional gauge theory
contains among the W-bosons also instanton particles\footnote{These
  become the usual instantons of the four-dimensional gauge theory
  once compactifying on $S^1$.} and the magnetic string. This instantonic particle is also called the 
  dyonic instanton, as it has both instanton number and electric charge \cite{Lambert:1999ua}. The magnetic string is
the uplift of the magnetic monopole in four-dimensions and has
therefore locally the same moduli space as its four-dimensional
companion. The worldvolume dynamics of this string is a $(0,4)$ CFT
with the magnetic monopole moduli space as a target. From the point of
view of geometric engineering, the string arises from wrapping four
dimensions of an M5-brane on the del Pezzo $\IB_{N_f}$. 

The next step is to compactify the five-dimensional theory on $T^2$
down to three-dimensional $\cN=4$ SYM \cite{arXiv:1107.2847}. This way
the magnetic string wrapped on $T^2$ becomes an instanton in the
three-dimensional theory. Computing such an instanton contribution
amounts to performing a certain path-integral over the instanton
moduli space. But in our case this is just the moduli space of
magnetic monopoles. On the other hand, from the viewpoint of the
M-theory setup the path integral is equivalent to computing the
partition function of the M5-brane, or, after compactifying on $T^2$
the partition function of (topologically twisted) $\cN=4$ SYM  on the
del Pezzo. This immediately opens the door to what was said in the
previous section. 

\subsection{Outline}

In Section 2 we start by describing the worldvolume theory of the
two-dimensional magnetic string from the viewpoint of five-dimensional
supersymmetric gauge theory. The metric on the moduli space captures the dynamics of
the magnetic string and shall be reviewed in some detail before
turning to the description of the action and conformal field
theory. In Section 3 we then proceed to a
presentation of the geometric engineering picture which contains a
description of the Coulomb branch in terms of moduli of the
Calabi-Yau. This way it is possible to make contact with the M5-brane
and the gauge theory on the del Pezzo surface. Finally, in Section
4 we formulate and provide tests of the conjecture. For magnetic charge $r=1$, we explicitly compute both elliptic genus and
the partition function of the  $\cN=4$ SYM on del Pezzo surfaces, and show they are the same.
Furthermore, we provide predictions for the elliptic genus of the CFT for $r= 2$ by computing the partition function on the
4d side. By compactifying the 5d gauge theory on a circle, we also make the connection to the BPS states of four-dimensional $\cN=2$ SYM providing further insight and evidence for the conjecture. We end the paper with a short conclusion and outlook in Section 5.

\section{$(0,4)$ CFT's from 5d supersymmetric gauge theories}
\label{sec:CFTfrom5d}

Throughout this section, we consider five-dimensional $\cN=1$ supersymmetric gauge theory with gauge group $G=SU(2)$ and $N_f\leq 8$ massless flavors in the fundamental representation \cite{Seiberg:1996bd}. The extension to massive hypermultiplets is considered at the end of this section. On the Coulomb branch of five-dimensional supersymmetric gauge theories, the spectrum contains a magnetic string as a solitonic BPS configuration. It can best be understood as the uplift of a BPS
magnetic monopole in four dimensions. The gauge group $SU(2)$ is broken to $U(1)$, and we denote the vacuum expectation value of the real adjoint scalar in five dimensions by $\phi$. Using a Weyl-reflection, we can always assume it to be positive. 

The tension $T$ of the BPS magnetic string is given by\footnote{Compared to \cite{arXiv:1107.2847}, our formula contains a factor of one-half. As $\frac{T}{\sqrt{2}} = r \phi_D = r \frac{\partial \cF}{\partial \phi}$, one can think of this as a rescaling of the prepotential $\cF$ by $\frac{1}{2}$. This is done to suppress factors of two in later sections. }
\begin{equation}
\label{eq:tension}
\frac{T}{\sqrt 2}=r\Big(\frac{\phi}{2 g_5^2}+\frac{\kappa}{4}\phi^2\Big)\ ,
\end{equation}
where $g_5$ is the bare (dimensionfull) five-dimensional Yang-Mills coupling constant and $\kappa$ determines the one-loop correction to the effective coupling constant. The factor $\kappa$ is also the coefficient in front of the one-loop induced Chern-Simons term in five dimensions \cite{Seiberg:1996bd}, which we assumed to be absent in the classical, microscopic theory. For $N_f$ massless flavors, we have
\begin{equation}
\kappa=2(8-N_f)\ .
\end{equation}
The requirement $N_f\leq 8$ alluded to in the beginning of this section, is reflected by the fact that the magnetic string must have positive tension (we always assumed $\phi>0$ without loss of generality). For $N_f>8$ this tension becomes negative for some values of $\phi$ and moreover, the metric on the Coulomb branch of the five-dimensional theory becomes singular \cite{Seiberg:1996bd}. Hence we do not consider $N_f>8$.

The spectrum also contains another solitonic BPS state, namely the dyonic instanton \cite{Lambert:1999ua}. It is the uplift of a four-dimensional instanton to a point particle in five dimensions and is therefore classified by an instanton charge $n_I$. For charge $n_I=1$, the dyonic instanton contributes to the mass, or central  charge\footnote{Again, compared to \cite{arXiv:1107.2847} , our central charge formula contains a factor of one-half. This is consistent with formula  (\ref{eq:tension}) as the instanton charge is given by $Z_I = \frac{\partial^2 \cF}{\partial \phi^2}$.}
\be
Z_I=\frac{1}{2g_5^2}+\half\kappa\,\phi\ ,
\ee
where the second term is due to the one-loop induced Chern-Simons term. The dyonic instanton 
also acquires an electric charge $n_e$ in five dimensions. The total central charge is then given by
\begin{equation}
\label{eq:Z}
Z=n_e\phi + n_I Z_I\ .
\end{equation}

In this section, we show how a
two-dimensional $(0,4)$ CFT emerges from the dynamics of collective
coordinates of the magnetic string. As we compactify a single-charged magnetic string on a circle, its momentum and winding modes are related to dyonic instanton charges $n_e$ and $n_I$, as we will show. This is similar to how a magnetic monopole can acquire electric charge and become dyonic. Furthermore we can consider the
magnetic string in five-dimensional supersymmetric gauge theory on
$\IR^3\times T^2$. When the worldsheet of the string wraps the torus
$T^2$, it manifests itself as an instanton in $\IR^3$. Such an
instanton corrects the metric on the Coulomb-branch of the
three-dimensional low-energy effective action through an instanton
induced four-fermion correlator, see
\cite{hep-th/9607163,Dorey:1997ij}.

\subsection{Magnetic string dynamics and quantization}

Before we discuss the dynamics of the magnetic string, we first review
some well-known aspects of magnetic monopole dynamics. They are BPS
objects in four-dimensional supersymmetric gauge theories, which we
think of as the zero radius limit of the five-dimensional gauge theory
on $\IR^4\times S^1$. For some background material on BPS magnetic monopoles, see
e.g. \cite{Tong:2005un,Weinberg:2006rq}. Our strategy to obtain the $(0,4)$ CFT, is to
lift the supersymmetric quantum mechanics of the monopole to a
two-dimensional sigma-model defined on the worldsheet of the magnetic
string. These sigma models have also been derived in \cite{Gauntlett:2000ks}
in an even more general setting, but without the link to 5d supersymmetric gauge theories. Moreover,
we are interested in the properties of the corresponding conformal field theory, as we discuss below.

Consider a static magnetic monopole of charge $r$ satisfying the
Bogomol'nyi equations on $\IR^3$. For gauge group $G=SU(2)$, such a
solution is parametrized by $4r$ bosonic collective coordinates on the
moduli space    
\begin{equation}
\cM_r=\IR^3 \times \frac{S^1\times {\widetilde \cM}_r}{\IZ_r}\ .
\end{equation}
For $r=1$, the moduli space is just the universal factor 
\begin{equation}\label{USM}
\cM_1=\IR^3\times S^1\ ,
\end{equation}
parametrized by the three positions of the monopole $\vec X\in \IR^3$
and a gauge orientation zero mode which is an angle denoted by
$\theta \in [0,2\pi]$.  

In general, both $\cM_r$ and ${\widetilde \cM}_r$ are hyperk\"ahler
manifolds, and e.g. ${\widetilde \cM}_2$ is (the double cover of) the
Atiyah-Hitchin manifold \cite{AH}. The dimension of the relative
moduli space ${\widetilde \cM}_r$, obtained after taking out the center of
mass coordinate, is $4(r-1)$. The discrete symmetry $\IZ_r$ acts on
both the $S^1$ as on ${\widetilde \cM}_r$. Its precise action is of not
much importance for our analysis, hence we refrain from giving its definition. 
The group of continuous isometries of the moduli space $\cM_r$ is the product of
Euclidean group in three dimensions and a phase rotation, 
\begin{equation}
G_{isom}=SO(3) \ltimes \IR^3 \times U(1)\ .
\end{equation}
The translation group $\IR^3$ simply acts as translations on the
center of mass coordinates $\vec X$. The rotations also act as
rotations on $\vec X$, but its action on ${\widetilde \cM}_r$ is more
complicated in general. For the Atiyah-Hitchin
manifold ${\widetilde \cM}_2$, it is known that there is an $SO(3)$ group of isometries which rotates the complex structures on ${\widetilde \cM}_2$ \cite{AH}. Finally, the $U(1)$ factor arises from the
large gauge transformations. On
the moduli space, it acts by rotation on the $S^1$. 

In the fermionic sector, we have $4r$ zero modes coming from the
adjoint fermions in the vector multiplet and $2rN_f$ from the fermions
in the hypermultiplets. These zero modes are solutions of the massless
three-dimensional Dirac equation in the presence of the magnetic
monopole, and their number can be computed using Callias index
theorems \cite{Callias:1977kg}. 
We parametrize the fermionic zero modes by Grassmann odd collective coordinates
\begin{equation}\label{GCC}
\psi^m\ ,\quad m=1,...,4r\ ,\qquad \chi^A\ ,\quad A=1,...,2rN_f\ .
\end{equation}
When masses for the hypermultiplets are present, the counting of the
fermionic zero modes changes. We discuss this in the next subsection. 

In the semiclassical approach of soliton quantization, one is
interested in the fluctuations around constant (static)
configurations, i.e. in the dynamics of the solitonic particle. To
study this,  we let all the collective coordinates and fermionic zero modes depend on time, and
study the motion of the monopole in the low-energy approximation
(i.e. for small velocities). This leads to a formulation in terms of a 
supersymmetric quantum mechanics on the moduli space with action
given by \cite{Sethi:1995zm,Cederwall:1995bc,Gauntlett:1995fu}, 
\begin{equation}\label{QM-act}
S_{QM}=\frac{1}{2}\int_{-\infty}^{+\infty}{\rm d}t\, \Big[g_{mn}\Big({\dot X}^m{\dot X}^n+i\psi^mD_t\psi^n\Big)+i\chi^AD_t \chi^A-\frac{1}{2}F_{mnAB}\,\psi^m\psi^n\chi^A\chi^B\Big]\ ,
\end{equation}
where $g_{mn}$ is the metric on the moduli space $\cM_r$ with
coordinates $X^m; m=1,...,4r$. The supersymmetric quantum mechanics defined by \eqref{QM-act} has
four supercharges. These are the four supercharges that are unbroken
by the BPS magnetic monopole. Four supercharges require the metric $g$
on $\cM_r$ to be hyperk\"ahler, which is known to be the case.
The covariant derivatives in \eqref{QM-act} are defined as 
\begin{equation}
D_t \psi^m={\dot \psi}^m+{\dot X}^n\Gamma^m{}_{np}\psi^p\ ,\qquad D_t \chi^A={\dot \chi}^A+{\dot X}^m\omega_{m\,B}^{A}\chi^B\ ,
\end{equation}
where $\Gamma$ is the Levi-Cevita connection on $\cM_r$ and $\omega$
is a connection on the $\cO(r)$ index bundle whose curvature is
$F_{mnAB}$ \cite{Manton:1993aa}.  

For $r=1$ the moduli space is given by the universal factor
\eqref{USM}. The action for the supersymmetric quantum mechanics for $r=1$ becomes
free, 
\begin{equation} \label{QMaction}
S_{QM}^{r=1}=\frac{1}{2}\int_{-\infty}^{+\infty}{\rm d}t\, \Big[g_{mn}\Big({\dot X}^m{\dot X}^n+i\psi^m {\dot \psi}^n\Big)+i\chi^A {\dot  \chi}^A\Big]\ ,
\end{equation}
with metric given by $g_{mn}{\dot X}^m{\dot
  X}^n=M|\dot{\vec{X}}|^2+(M/|\phi|^2)\,\dot{\theta}^2$, where $\phi$
is the vev of the scalar field, in the notation of
\cite{arXiv:1107.2847}, and $M$ is the monopole mass. Finally,
$\theta\in [0,2\pi]$ is a global $U(1)$ charge angle that parametrizes
the circle $S^1$ in the moduli space. The radius of this circle is given by $R=1/\phi$ in our conventions. The gauge transformation
$-\mathrm{diag}(1,1)\in SU(2)$ leaves invariant the gauge and Higgs field, but
acts on the fermionic zero modes by $-1$. Therefore, the $O(1)=\{1,-1\}$ index bundle
is the M\"obius bundle \cite{Manton:1993aa}, i.e.  the flavor fermions change sign after making a half-period around 
the $S^1$:
\be \label{Mbundle}
\theta \to \theta +\pi, \qquad \chi^A \to -\chi^A.
\ee
This transformation must leave invariant the spectrum of the supersymmetric quantum mechanics.
Leaving out for notational simplicity the fermions $\psi^m$, which are invariant under \eqref{Mbundle}, then a general ground-state in the quantized Hilbert space of the quantum mechanics (\ref{QMaction}) is given by
\begin{equation}
	|\Psi\rangle = e^{i n_e \theta} e^{i \vec{k} \cdot \vec{X}} \otimes \chi^{A_1} \cdots \chi^{A_n}|0\rangle,
\end{equation}
where $\vec{k}$ is the momentum along the non-compact directions $\vec{X} = \{X^1,X^2,X^3\}$ and $n_e$ is the momentum along the compact $S^1$. A nonzero momentum along $S^1$ makes the monopole dyonic with electric charge $n_e$. Furthermore, the $\chi^A$ are viewed as operators satisfying the Clifford-algebra
\begin{equation}
	\{\chi^A, \chi^B\} = \delta^{AB}.
\end{equation}
Under the transformation (\ref{Mbundle}) this state changes as follows
\begin{equation}
	|\Psi\rangle \mapsto e^{i n_e (\theta + \pi)} e^{i \vec{k} \cdot \vec{X}} \otimes (-\chi^{A_1}) \cdots (-\chi^{A_n})|0\rangle = (-)^{n_e} (-)^H |\Psi\rangle,
\end{equation}
 where use has been made of the fact that the chirality operator
 $(-)^H$ anti-commutes with the $\chi^A$ for $A=1, \cdots, 2N_f$. As
 (\ref{Mbundle}) must be a symmetry of the quantum mechanics we arrive
 at the constraint \cite{Seiberg:1994aj,Gauntlett:1995fu} 
 \begin{equation}\label{QMconstr}
	(-)^{n_e} |\Psi\rangle = (-)^{H} |\Psi\rangle.
\end{equation}
This constraint implies a correlation between the electric-magnetic and flavor charges
of the four-dimensional gauge theory, which avoids the unphysical
decay of the BPS particles \cite{Seiberg:1994aj}, for example decay of
the W-boson into two particles of electric charge one.
 
It is now easy to uplift the quantum mechanics to a two-dimensional sigma model. The monopole becomes a string in five dimensions, and the supersymmetric quantum mechanics becomes a non-linear two-dimensional sigma model. The uplift is done in such a way that it preserves four supercharges, 
since the magnetic string is BPS in five dimensions. Therefore, its worldsheet dynamics must preserve either
 $(2,2)$ or $(0,4)$ supersymmetry. However, the multiplet structure of
 such models imposes strong constraints, and only $(0,4)$
 supersymmetry is possible. The reason is that the $\chi^A$ do not
 have bosonic superpartners, whereas $\psi^m$ sit in the same
 multiplet as the bosonic coordinates $X^m$. As a consequence, the
 $\chi^A$ must be part of the left-moving sector, and $\psi^m$ sits in
 the right-moving supersymmetric sector. 

The action for the $(0,4)$ model is, 
\begin{equation}\label{CFT}
S={\tilde T}\int_{T^2}{\rm d}^2\sigma \Big[g_{mn}\Big({\partial_+ X}^m{\partial_- X}^n+i\psi^mD_+\psi^n\Big)+i\chi^AD_- \chi^A-\frac{1}{2}F_{mnAB}\,\psi^m\psi^n\chi^A\chi^B\Big]\ ,
\end{equation}
where we introduced coordinates on the worldsheet $\sigma^{\pm}$ and defined
\begin{equation}
D_+ \psi^m={\partial_+ \psi}^m+{\partial_+ X}^n\Gamma^m{}_{np}\psi^p\ ,\qquad D_- \chi^A={\partial_- \chi}^A+{\partial_- X}^m\omega_{m\,B}^{A}\chi^B\ .
\end{equation}
The tension ${\tilde T}$ will be proportional to the tension $T$ in \eqref{eq:tension} by a numerical factor that we determine below. For later purposes, to define the elliptic genus, we have defined the $(0,4)$ model on a two-torus $T^2$. The magnetic string then wraps the $T^2$ and becomes an instanton in the three-dimensional gauge theory.

The action \eqref{CFT} falls into the class of $(0,4)$ supersymmetric sigma models \cite{Howe:1988cj,Hull:1993ct,Papadopoulos:1993mf}. The most general Lagrangian with $(0,4)$ also contains mass terms and a $b$-field, but they do not appear in our setup. In that case, $(0,4)$ supersymmetry implies the target space to be a hyperk\"ahler manifold $({\cal M},g)$ equipped with a vector bundle $E$ with connection $\omega$.  In the absence of masses or scalar potential on $\cal M$, the action defines a conformal field theory. It was argued in \cite{Howe:1992tg} that conformality can be maintained at the quantum level\footnote{See also \cite{Witten:1994tz,Lambert:1995dp} for related discussions. If the $(0,4)$ sigma model we consider is not conformal at the quantum level, then we assume that it flows to some fixed point in the infrared. The cases we discuss explicitly in this paper, namely the ones with $r=1$ are (orbifolds of) free field theories and hence are true CFT's.} . The central charges in the left and right-moving sectors can be computed in the limit of zero coupling constant, where the interaction between the monopoles vanishes. Using this argument, the calculation of the central charges is that of free fields for which we obtain
\begin{equation}
c_L=r(4+N_f)\ ,\qquad c_R=6r\ .
\end{equation}
For magnetic charge $r=1$, the $(0,4)$ model becomes free, since the curvature of the $O(1)$ bundle vanishes. The resulting action is that of a free $(0,4)$ CFT with action
\begin{equation}\label{freeCFT}
S={\tilde T}\int_{T^2}{\rm d}^2\sigma \Big[g_{mn}\Big({\partial_+ X}^m{\partial_- X}^n+i\psi^m\partial_+\psi^n\Big)+i\chi^A\partial_- \chi^A\Big]\ ,
\end{equation}
containing three non-compact bosons $\vec X$, one compact boson $\theta$. In the right-moving sector, there are four fermionic superpartners, and in the left-moving sector there is
a flavor symmetry group $SO(2N_f)$ rotating the fermions $\chi^A; A=1,...,2N_f$. The metric components $g_{mn}$ are constant in this basis and given in the text after \eqref{QMaction}.

Upon uplifting the quantum mechanics for the monopole to the 2d CFT of
the magnetic string one has to take into account the fundamental
constraint (\ref{QMconstr}). This is done as follows. On the one hand,
one can see from the mode expansion of the left-moving fermions  
\begin{equation}
	\chi^A = \sum_{n} \rho_n^A e^{-n (t+i x)}\ ,
\end{equation}
that the action $\chi^A \mapsto - \chi^A$ translates to
\begin{equation}
	\rho_n^A \mapsto - \rho_n^A\ ,
\end{equation}
which induces the natural generalization of the chirality-operator $H$ to the fermion number operator $(-)^F$ acting on physical states. Now states in the Hilbert space of the magnetic string have to satisfy the constraint
\begin{equation}\label{cftconstr}
	(-)^{n_e + F} |\theta\rangle \otimes |\chi \rangle = |\theta\rangle \otimes |\chi\rangle,
\end{equation}
where $|\theta\rangle$ denotes a state in the Hilbert space of the compact boson $\theta$ and $|\chi\rangle $ is obtained by quantizing the fermions $\chi^A$. But this is nothing else than an orbifold of a 2d CFT by a group action $G=\{1,g\}$ generated by the identity and $g=(-)^{n_e + F}$.

On the other hand, unlike in quantum mechanics, the momentum of the magnetic string along $S^1$ is no longer simply identified with the electric charge $n_e$. The identification of the electric charge is rather deduced from the excitation modes of the magnetic string and their relation to BPS states of the gauge theory. This works as follows. For the left- and right-moving Hamiltonians we have the decomposition
\begin{equation}
L_0 = \frac{1}{2} p_L^2  + N, \quad \overline{L}_0 = \frac{1}{2} p_R^2 + \overline{N},
\end{equation}
where $N$ and $\overline{N}$ denote the remaining left- and
right-moving oscillator modes (including those of the $2N_f$
left-moving fermions), and
\begin{equation} \label{eq:pRL}
(p_L,p_R)=\frac{1}{{\sqrt {\tilde T}}}\Big(p\,\frac{\phi}{2}-w\,\frac{\tilde T}{\phi},\,
p\,\frac{\phi}{2}+w\,\frac{\tilde T}{\phi}\Big)\ ,
\end{equation}
are winding and momentum excitations along the target space $S^1$ satisfying the usual relation
\begin{equation}
p_R^2-p_L^2=2\,p\,w\ .
\end{equation}
The $(0,4)$ elliptic genus projects on the right-moving
sector to states which preserve four fermionic operators of the large $\cN=4$
algebra \cite{Sevrin:1988ew, Maldacena:1997de, Manschot:2008}:
\be
\label{eq:barL}
\overline{L}_0-c_R/24=\frac{\left(p\, \frac{\phi}{2} + w\, \frac{ \tilde{T}}{\phi}\right)^2}{ 2\tilde{T}}\ .
\ee

We continue now with comparing these quantum numbers with those of the compactified gauge theory
on $\mathbb{R}^3\times T^2$. For simplicity we let the torus be
the direct product $S^1_1\times S^1_2$, where $S^1_1$ is the Euclidean
time circle of radius $R_1$ and $S^1_2$ is the spatial circle wrapped
by the magnetic string with radius $R_2$. Therefore, the volume of the
torus is given by $V_{T^2}=R_1R_2$ and the complex structure modulus by
$\tau=i\frac{R_1}{R_2}=i\tau_2$. 

The 5-dimensional mass of the magnetic string carrying electric and
instanton charge takes the form \cite{Chen:2010yr}:
\be
M_{\mathrm{5d}}=\left|R_2\frac{T}{\sqrt{2}}+\frac{N'}{R_2}+i(n_e \phi + n_I Z_I)\right|,
\ee
where $N'$ is the momentum around the spatial $S^1_2$.
We consider the limit $R_2\,T\gg \phi, Z_I,\frac{1}{R_2}$. In this
limit, the mass has the following expansion\footnote{A very similar
  expansion for the mass of D4/M5-branes was obtained in \cite{Denef:2007vg}.}
\be
\label{eq:M5d}
M_\mathrm{5d}=R_2\frac{T}{\sqrt{2}}+\frac{(n_e \phi + n_I Z_I)^2}{\sqrt{2}\,R_2\,T}+\frac{N'}{R_2}+\dots,
\ee
where the dots represent terms with increasing negative powers in $T$, which vanish in the limit. The energy of the excitations
of the CFT provide the corrections to the leading term
$R_2\,T/\sqrt{2}$. This gives for the dimensionless action $R_1\,
M_\mathrm{5d}$ in terms of the CFT operators:
\be
R_1\, M_\mathrm{5d}=V_{T_2}\frac{T}{\sqrt{2}}+\tau_2\left(L_0-c_L/24 +\overline{L}_0-c_R/24\right).
\ee 
 Using Eqs. (\ref{eq:Z}), (\ref{eq:pRL}) and (\ref{eq:barL}) one finds for $r=1$
the following identification of CFT and 5d parameters:
\begin{equation} \label{wpbasis}
	w = n_I\ , \qquad p = n_e + \kappa\, n_I/4\ , \qquad  N=N'+pw+c_L/24\ ,\qquad {\tilde T}=\frac{T}{2{\sqrt 2}}\ ,
\end{equation}
that is, the winding number of the magnetic string corresponds to
instanton charge and the momentum corresponds to electric charge
shifted by a half-integral multiple of instanton charge.  
 
In conclusion, we have derived that the 2d $(0,4)$ CFT for $r=1$ and $N_f\leq 8$ flavors is an orbifolded free CFT. The winding and momentum charges in this CFT are related to the instanton and electric charges according to \eqref{wpbasis}. These facts will be important when we give a proof of our conjecture in Section 4.

\subsection{Massive hypermultiplets}
\label{sec:masshyp}

There are some important modifications in the analysis of the previous subsection that occur when the hypermultiplets are taken to be massive. We now discuss some of these effects. 

In the three-dimensional gauge theory, after compactification of the 5d theory on a torus $T^2$, we have $2N_f$ (two-component) Dirac species in the fundamental representation, which we denote by 
$\chi^i, {\tilde \chi}^i; i=1,...,N_f$. In the five dimensional theory, they just become $N_f$ Dirac fermions or, equivalently, $2N_f$ symplectic Majorana fermions on which a global flavor symmetry group $SO(2N_f)$ acts.

In three dimensions, one can add a triplet of mass terms for each species. First there is a complex mass that can be interpreted as coming from a superpotential in four dimensions. The complex mass terms for the fermions are of the form $M_i\chi^i{\tilde \chi}^i+c.c.$, with $M_i$ complex. Secondly, there is a real mass term, a Dirac mass, which looks like
\begin{equation}\label{5dmass}
{\cal L}_{mass}=-m_i( {\bar\chi}^i \chi^i + {\bar {\tilde \chi}}^i{\tilde \chi}^i)\ .
\end{equation}
Similar mass terms can be written for the hypermultiplet scalars in such a way that the theory remains supersymmetric. In fact, there is a simple way as how the real mass terms arise from a five dimensional point of view. Before adding masses, one can gauge a subgroup of the flavor group $U(1)^{N_f}\subset SO(2N_f)$. To do this, one introduces momentarily $N_f$ five-dimensional vector multiplets, each containing a real scalar field. One then freezes these vector multiplets, by putting them in a background in which the real scalars are non-zero and constant, with a vacuum expectation value equal to the masses $m_i$. On top of this, we can allow for non-trivial Wilson lines for the flavor gauge fields, after compactification on the torus with complex structure $\tau$. Hence we define
\begin{equation}
y_i\equiv {\rm Re}(y_i) + \tau\, {\rm Im}(y_i)=2\oint_{S^1_1} A_i+2\tau \oint_{S^1_2} A_i\ .
\end{equation}
This procedure preserves supersymmetry and generates mass terms as in \eqref{5dmass}.

Since only real mass terms have a five-dimensional interpretation, we set possible complex masses in three dimensions to zero. Combined with the relevant Yukawa terms for the hypermultiplet fermions, we get terms in the three-dimensional action of the form
\begin{equation}
{\cal L}_f= {\bar\chi}^i (\sigma-m_i)\chi^i + {\bar {\tilde \chi}}^i(\sigma - m_i) {\tilde \chi}^i\ .
\end{equation}
Here, $\sigma$ is the adjoint scalar field, whose vacuum expectation value we have denoted by
\begin{equation}
\sigma = \phi\, \tau_3\ ,
\end{equation}
and without loss of generality we can take $\phi>0$, as before.

The structure of the zero modes for the hypermultiplet fermions now changes. It follows from index theorems \cite{Callias:1977kg} that, for a given value of the flavor index $i$  (see also \cite{Dorey:1998kq}):
\begin{eqnarray}\label{jump}
\phi > |m_i| \qquad &\rightarrow& \qquad {\mbox  {$2r$ fermionic zero modes}}\ ,\nonumber\\
\phi < |m_i| \qquad &\rightarrow& \qquad {\mbox {no fermionic zero modes}}\ ,
\end{eqnarray}
and when $\phi=m_i$ for a particular value of $i$, the zero mode is still present but becomes non-normalizable, i.e. it is not square-integrable over $\IR^3$. So, only when $\phi > |m_i|$ for all values of $i$, we obtain the same number of fermionic collective coordinates as in \eqref{GCC}. 

One can include the effect of the mass terms on the $(0,4)$ sigma model. In fact, this was done in \cite{Gauntlett:2000ks}. However, we will not study the corresponding CFT in this paper and leave this as a topic for future research. The only relevant piece of information for the present discussion, is the jump in the number of fermionic zero modes given by \eqref{jump}. As we discuss in the next section, this jump can be interpreted in terms of geometric transitions when embedding the theory in M-theory, indicating that the 5d/2d/4d correspondence also works for massive hypermultiplets.

\section{M-theory embedding and $\cN=4, d=4$ SYM on $\IB_n$}
\label{sec:Mtheory}

The five-dimensional gauge theory of Section 2 can be embedded into
M-theory using either a brane construction \cite{hep-th/9710116} or in
terms of geometric engineering \cite{Douglas:1996xp, Morrison:1996xf,
  Intriligator:1997pq}. For the 4d/2d correspondence we are proposing,
the latter construction is more important and we shall follow it in
this section. The strategy here will be to identify the magnetic
string with the M5-brane wrapped around a del Pezzo surface and then
look at two different limits of the M5-brane world-volume theory. 
To obtain a $\cN=1$ gauge theory on $\IR^3 \times T^2$ we
compactify M-theory on $\textrm{CY}_3 \times T^2$ where
$\textrm{CY}_3$ is a local Calabi-Yau threefold of the following form 
\begin{equation}
	\begin{array}{ccc}
		\cO(K_{IB_{N_f}}) & \longrightarrow & \textrm{CY}_3 \\
		~        &  ~                 & \downarrow \\
		~        &  ~                 & \IB_{N_f}
	\end{array}
\end{equation}
In the above $\cO(K_{IB_{N_f}})$ is the canonical bundle over the surface
$\IB_{N_f}$ which itself is the Hirzebruch surface
$\IF_0=\IP^1 \times \IP^1$ blown-up  at $N_f$ points. 
 As these surfaces have positive anti-canonical class $- K_{\IB_{N_f}}>0$, they are rigid inside the Calabi-Yau. 
 
\subsection{The surface $\IB_n$ and lattice $H_2(\IB_n,\mathbb{Z})$}

\label{subsec:surfacelattice}
In this subsection, we explain in more detail aspects of the lattice
$H_2(\IB_n,\mathbb{Z})$ and interpret them in terms of the five
dimensional field theory following closely the discussion in \cite{Douglas:1996xp}. Five dimensional $\cN = 1$ $SU(2)$ gauge theory
is parametrized on the Coulomb branch by the bare coupling constant $g_5$, the vev
$\phi$ of the vector multiplet scalar, and the bare masses $m_i$, $i=1, \cdots, N_f$. 
All these parameters are related to the geometry of the del Pezzo surface as discussed below.

The homology group of $\IB_{n}$, $H_2(\IB_{n},\IZ)$, is generated by
the classes $\bff$ and $\bC$ corresponding to the two $\IP^1$'s of
$\IF_0$ (the first being the fiber $\IP^1_f$ and the other the base denoted
by $\IP^1_B$) and $n$ blow-up classes $\bc_i,~i=1, \cdots, n$. As
$\IP^1_f$ and $\IP^1_B$ have zero self-intersection and meet each
other exactly at one point transversely, we have 
\begin{equation}
	\bC^2 = \bff^2 = 0, \quad \bC \cdot \bff = 1\ .
\end{equation}
Furthermore, all other classes have self-intersection number $-1$ and
are mutually orthogonal to each other and to the two classes $\bC$ and
$\bff$. Thus we obtain the following intersection matrix 
\be
\label{eq:Qblowup}
\left( \begin{array}{ccccc} 0 & 1 & & & \\ 1 & 0 &  & & \\ & & -1 & &\\ & &
    & \dots & \\ & & & & -1\end{array} \right)\ .
\ee
In this basis the canonical class of the del Pezzo is given by 
\begin{equation} \label{cc}
	- K_{\IB_{n}} = 2 \bC + 2 \bff - \sum_{i=1}^{n} \bc_i\ .
\end{equation}

Geometrically, the del Pezzo surface $\IB_{n}$ can be viewed as a fiber space of $\IP_B^1$ where the generic fiber is $\IP_f^1$. The sizes of base and fiber $\IP^1$ are related to the gauge theory parameters as follows \cite{Morrison:1996xf}
\begin{equation}
\label{eq:JfJC}
	J \cdot \bff = 2 \phi\ , \quad J \cdot \bC = \frac{1}{4 g_5^2} + 2 \phi\ ,
\end{equation}
where $J$ is the pullback of the Calabi-Yau K\"ahler form to the del
Pezzo. In the weak coupling limit the volume of the base becomes
proportional to $1/g_5^2$. Moreover, over $n$ points the fiber has two
$\IP^1$'s intersecting at a point, i.e. the geometry of a resolved
$A_2$ singularity. Using the notation of \cite{Douglas:1996xp}, we denote each of the blow-up modes for the
$\IP^1$'s by $[A_2^{i,1}]$ and $[A_2^{i,2}]$. One arrives then at the relation: 
\begin{equation}
	\bff = [A_2^{i,1}] + [A_2^{i,2}]\ .
\end{equation}
This relation of the two classes with $\bff$ and the conditions
\begin{equation} 
	[A_2^{i,1}] \cdot [A_2^{i,2}] = 1, \quad [A_2^{i,1}]^2=[A_2^{i,2}]^2 = -1\ ,
\end{equation}
uniquely fixes these as follows
\begin{equation}
	[A_2^{i,1}] = \bff - \bc_i, \quad [A_2^{i,2}] = \bc_i\ .
\end{equation}

The volumes of these classes are related to the $SU(2)$ gauge theory parameters as follows
\cite{Morrison:1996xf, Douglas:1996xp}
\begin{equation}
\label{eq:hypermass}
	\quad J \cdot [A_2^{i,1}] = \phi + m_i\ , \quad J \cdot [A_2^{i,2}] = \phi - m_i\ , 
\end{equation}
where $m_i$, $i=1, \dots, n=N_f$ are hypermultiplet
masses. Combination of (\ref{eq:JfJC}) and (\ref{eq:hypermass}) gives for $J$:
\be
\label{eq:Jfieldtheory1}
J=\frac{\bff}{4\,g_5^2}-\phi\,K_{\IB_n}-{1\over 2}
\sum_{i=1}^{N_f} m_i (\bff-2 \bc_i).
\ee
The condition for $J$ to lie in the K\"ahler cone is:
\begin{eqnarray}
	0 & < & 2 \phi\ , \nonumber \\
	0 & < & \phi + m_i\ , \\
	0 & < & \phi - m_i \ . \nonumber 
\end{eqnarray}
The curves $[A_2^{i,1}]$ or $[A_2^{i,2}]$ can be flopped to give rise
to other birational models of the Calabi-Yau $X$. More precisely, when
$\phi = m_i$ one of the $\IP^1$'s $[A_2^{i,2}]$ has shrunk to zero
size. At this point in moduli space we can do a flop in the Calabi-Yau
eliminating this $\IP^1$ in the del Pezzo and growing another $\IP^1$
in the Calabi-Yau orthogonal to the del Pezzo. In this process the del
Pezzo $\IB_n$ gets replaced by $\IB_{n-1}$. 	Notice that this is in
complete agreement with the 2d side, as one of the flavor fermionic
zero modes becomes non-renormalizable and hence disappears from the
CFT. 

The exceptional divisors of the blow-up maps $\phi_i:\IB_i\to
\IB_{i-1}$ together with the generic fiber class $\bff$ and the base
class $\bC$ span as lattice ellements $(\bC,\bff,\{\bc_i\})$ the
unimodular lattice $\Lambda_{\IB_n}\cong H^2(\IB_n,\mathbb{Z})$ whose
quadratic form is given by (\ref{eq:Qblowup}). In the following, we
will abbreviate $\Lambda_{\IB_n}$ to $\Lambda$. 
The signature of $\Lambda$ is thus $(1,b_2-1)=(1,n+1)$. We let
$G(\Lambda)$ be the Grassmannian of positive definite subspaces of
$\Lambda \otimes \mathbb{R}$, this space is $b_2-1$ dimensional.
A choice of $J\in \Lambda\otimes \mathbb{R}$ such that $J^2>0$
corresponds to a point in $G(\Lambda)$, and determines a split of
$\Lambda$ into positive and negative definite lattices denoted
by $\Lambda_{\pm}$.  The projection of a vector $\bfk\in \Lambda$ to $\Lambda_{+}$ is given by
\be
\bfk_+=\frac{\bfk\cdot J}{J^2}\,J\ ,
\ee
and to $\Lambda_{-}$ by $\bfk_-=\bfk-\bfk_+$. In this paper we will be
mainly dealing with the situation where all mass-parameters $m_i,
~i=1, \cdots, N_f$ are zero. In this case the 5d gauge theory exhibits
an enhanced $SO(2N_f)$ global flavor symmetry.  
As was already noted in section 2 this symmetry also shows up in the
left-moving fermionic sector of the magnetic string. In order to make
this symmetry more manifest we will perform a basis change of the
lattice $\Lambda$. We want in particular to extract a $D_n$
lattice. 
To this end we write $\Lambda$ as a direct sum: $A\oplus
D$ with the lattice $A$ spanned by
$\bfa_1=-K_{\IB_n}$ and $\bfa_2=\bff$, and the lattice $D$ spanned by the remaining $n$ directions denoted
by $\bd_i$.  One obtains for the quadratic form in the new basis:
\be
\label{eq:Amatrix}
\left( \begin{array}{ccc} 8-n & 2 &  \\ 2 & 0 &   \\ & &
    -\mathcal{Q}_{D_n} \end{array} \right)\ ,
\ee
where $\mathcal{Q}_{D_n}$ is the $D_n$ Cartan matrix. We will denote the projection of a vector $\bfk\in \Lambda$ to the lattice $A$
(respectively $D$) by $\bfk_A$, respectively $\bfk_D$. 

Thus the uni-modular lattice $\Lambda$ can be represented as a gluing of the
two lattices $A$ and $D$. Lattice points of $\Lambda$ can be written
as $\bsa+\bd$, with in general $\bsa\in A^*$ and $\bd \in D^*$. We are interested in
representatives for the coset $A^*/A$ and $D^*/D$, which are called
``glue vectors'' of $A$ respectively $D$ \cite{Conway:1999}.   
Obtaining the uni-modular lattice $\Lambda$ from $A$ and $D$ amounts
to a choice of isomorphism between $A^*/A$ and $D^*/D$. Since $\det
A=\det D=4$ this isomorphism is given by four vectors 
$\bg_i=(\boldsymbol a_i,\boldsymbol d_i)$, which we will also refer to as
gluing vectors. The glue vectors of $A$ can be found by acting
with the inverse $\cQ_A^{-1}=\frac{1}{4}\left(\begin{array}{cc} 0 & 2\\ 2 & n-8 \end{array}\right)$ on $(0,0)$, $(1,0)$, $(0,1)$ and
$(1,1)$, and similarly for those of $D$.\footnote{The case $n=0$ needs to be treated separately;
since the determinant of the quadratic form is 4 in the new basis,
only one vector in $A^*/A$ needs to be
added to recover the original lattice. Equivalently one can
choose as basis vectors $\bfa_1=-\half K_{\IB_0}$ and $\bfa_2=\bff$.} These
gluing vectors correspond to correlation of charges in the spectra of
the field theories, for example between electric and flavor charge \cite{Seiberg:1994aj, Gauntlett:1995fu}, as will be explained in more detail
in Section \ref{sec:4dimensions}.  
  
Most qualitative aspects can be understood by taking $n=1$. The
lattice $D$ is generated by $\bfd_1=-\bff+2\bc_1$. For $A\oplus D$, the gluing vectors are:
\begin{eqnarray}
\label{eq:gluingn=1}
\bg_0&=&{\bf 0},\,\qquad \qquad \qquad\, \bg_1=\textstyle{1\over 4}(2,-5,3),\\
\bg_2&=&\textstyle{1\over 4} (0,2,2),\,\quad\qquad \bg_3=\textstyle{1\over 4}(2,-7,1).\non
\end{eqnarray}
We observe in particular that $\exp\left(2\pi i\sum_{j=1}^3 \bg_{i,j}\right)=1$, and also that $\exp\left(4\pi i\sum_{j=2}^3 \bg_{i,j}\right)=1$. 
We choose as generators for $D$ when $n=2$:
\begin{eqnarray}
\bfd_1=-\bff+\bc_1+\bc_2,\qquad & \bfd_2 =-\bff+2\bc_2,
\end{eqnarray}
with gluing vectors:
\begin{eqnarray}
\bg_0&=&{\bf 0},\,\qquad \qquad \qquad \, \bg_1=\textstyle{1\over 2}(1,0,1,1),\\
\bg_2&=&\textstyle{1\over 2} (0,1,0,1),\,\quad\,\quad \bg_3=\textstyle{1\over 2}(1,-1,1,0).\non
\end{eqnarray}
Similarly for $n=3$ we have:
\be
\bfd_1=-\bff+\bc_1+\bc_2, \qquad \bfd_2=-\bff+\bc_2+\bc_3, \qquad \bfd_3=-\bff+\bc_1+\bc_3,
\ee
and
\begin{eqnarray}
\label{eq:gluingn=3}
\bg_0&=&{\bf 0},\,\qquad \qquad\qquad \qquad \quad \,\,\, \bg_1=\textstyle{1\over 4}(2,-3,-1,-1,3),\\
\bg_2&=&\textstyle{1\over 4} (0,2,-2,2,2),\,\quad\qquad \,\,\,\,\bg_3=\textstyle{1\over 4}(2,-5,1,1,1).\non
\end{eqnarray}
A natural expansion of the K\"ahler modulus $J$ in terms of this basis
is:  
\be
\label{eq:Jfieldtheory}
J=\frac{\bff}{4\,g_5^2}-\phi\,K_{\IB_n}+2\sum_{i,j=1}^{N_f} \tilde m_i\, Q_{D_{N_f},ij}^{-1}\, \bd_j,
\ee
where $Q_{D_{N_f},ij}^{-1}$ is the inverse of the matrix
$Q_{D_{N_f},ij}=-\bd_i\cdot\bd_j$ in (\ref{eq:Amatrix}). For $N_f=1$, we have $\tilde m_1=m_1$, and
for $N_f=2,3$, the $\tilde m_i$ can be expressed as simple linear
combinations of the $m_i$. This finishes our discussion of the basis change (\ref{eq:Amatrix}).

We briefly mention a representation of the lattice $\Lambda$
as a gluing of the 1-dimensional lattice spanned by $K_{\IB_n}$ and the $E_{n+1}$-lattice,
where the quadratic form of the $E_{n+1}$-lattice is given by the Cartan
matrices of $E_1=SU(2)$, $E_2=SU(2)\times U(1)$, $E_3=SU(3)\times SU(2)$, $E_4=SU(5)$,
$E_5=Spin(10)$, and for $n=6,7,8$ $E_n$ are the usual exceptional groups.
Bases for $E_n$ are given by: 
\begin{eqnarray}
\label{eq:basisEn}
&&\, \bfe_1= \bC-\bff,  \qquad n=0, \non \\
&& \left. \begin{array}{l} \bfe_1= \bC-\bff,  \\
    \bfe_2=\bc_1 \end{array} \right\}  \qquad n=1, \\
&& \left. \begin{array}{l} \bfe_1=\bC-\bff \\ \bfe_2 =\bff-\bc_1-\bc_2\\
    \bfe_i=\bc_{i-2}-\bc_{i-1}, \quad i=3,\cdots, n+1 \end{array} \right\} \qquad n \geq 2\non       
\end{eqnarray}
This basis becomes important in the infinite coupling limit $g_5
\rightarrow \infty$ and captures the exceptional symmetry of the
non-trivial fixed points  these theories \cite{Morrison:1996xf, Seiberg:1996bd}.

\subsection{Magnetic strings from M5/$\IB_n$}

Having identified the parameters of the field theory in terms of
moduli of the del Pezzo we now turn to the identification of
states. Here we note from \cite{arXiv:1107.2847} that the
W-bosons correspond to M2-branes wrapped around the $\IP^1_f$ and
states with topological instanton charge correspond to wrapping around
$\IP^1_B$. Furthermore, the M5-brane wrapped around the whole del
Pezzo corresponds in the 5d gauge theory to the magnetic string with
tension 
\begin{equation}
	\frac{T}{\sqrt{2}} =Z_m= \textrm{Vol}(\IB_{N_f}) = \frac{1}{2} \int_{\textrm{CY}_3} [\IB_{N_f}] \wedge J \wedge J\ ,
\end{equation}
as was also shown in \cite{arXiv:1107.2847}. Apart from magnetic charge $n_m=r$, which corresponds to five-brane
wrappings, a generic state can have electric $n_e$, instanton $n_I$ and flavor
charge $n_{f,i}$, and is parametrized by an element of the second homology
lattice as follows 
\be
\label{eq:chargevector}
	\half n_e \bff - n_I K_{\IB_{N_f}}-\half \sum_{i=1}^{N_f} n_{f,i} \,\bd_i \in H^2(B_{N_f},\mathbb{Z})+rK_{\IB_{N_f}}/2\ .
\ee
The last term is due to the non-integral flux-quantization on the M5-brane \cite{Witten:1996hc}.
The new ingredients here are the fundamental matter particles which can be identified with M2-branes wrapping the curves $A_2^{i,1}$ as well as $A_2^{i,2}$ with a different orientation. Thus, to summarize, we arrive at Table 1 which gives us the dictionary between field theory and M-theory.

\begin{table}[h]
\begin{center}
\begin{tabular}{|c|c|c|}
    \hline
    ~                       & field theory       & M-theory     \\
    \hline
    coupling                & $\frac{1}{g_{5,0}^2}$  & $\phi_B$     \\
    moduli                  & $\phi,~m_i$             & $\phi_f,~m_i$     \\
    \hline
    \multirow{4}{*}{states} & W-bosons           & $M2/\IP^1_f$ \\
                            & fundamental matter       & $M2/\{A_2^{i,1}-A_2^{i,2}\}$ \\
                            & dyonic instantons  & $M2/\{n \IP^1_B + m \IP^1_f\}$ \\
                            & magnetic string            & $M5/\IB_{N_f}$   \\
    \hline
\end{tabular}
\end{center}
\caption{Dictionary between five-dimensional field theory and M-theory. In the above $n$ and $m$ are arbitrary integers and refer to the number of wrappings around $\IP^1_B$ and $\IP^1_f$.}
\end{table}

We now claim that in the regime where the characteristic lengths
$\ell_{T^2}$ and $\ell_{\IB_{N_f}}$ of $T^2$ and $\IB_{N_f}$ are such that:
\begin{equation}
	\ell_{T^2} \gg \ell_{\IB_{N_f}},
\end{equation}
the dynamics of $r$ M5-branes wrapped around $\IB_{N_f}$ is governed
by the $(0,4)$ sigma model on $T^2$, whose target space we propose to
be equal to that of $r$ $SU(2)$ monopoles introduced in
Section 2. The chirality of the sigma model can be derived from the
chirality of the $(0,2)$ theory of $r$
M5-branes. Furthermore, the number of supercharges can be deduced
consistently from two viewpoints. First of all, as the M5-brane is the
solitonic string in 5d $\cN=1$ SYM, its worldvolume preserves half of
the supersymmetry giving us 4 supercharges. On the other hand, the
topological twist on the del Pezzo surface preserves exactly 4
supercharges due to an argument in \cite{Vafa:1994tf}.

\subsection{$\cN = 4, d = 4$ SYM on $\IB_n$ from M5/$T^2$}
\label{subsec:N4YM}

Compactifying the worldvolume theory of $r$ coincident M5-branes on
$T^2\times \IB_{N_f}$ in the regime
\begin{equation}
	\ell_{T^2} \ll \ell_{\IB_{N_f}}\	,
\end{equation}
will give rise to topologically twisted $\cN = 4$ $U(r)$ Yang-Mills theory on
$\IB_{N_f}$ \cite{Minahan:1998vr} studied in detail in Ref. \cite{Vafa:1994tf}. In this setup the complex structure $\tau$ of $T^2$ gets identified with the complexified $U(r)$ gauge coupling 
\begin{equation}
	\tau = \frac{4\pi i}{g^2} + \frac{\theta}{2\pi}\ .
\end{equation}
The $SL(2,\IZ)$ symmetry of the $T^2$ then descends to S-duality acting on the gauge
coupling, implying that the partition function of the topological
theory transforms as a modular form. In fact, the partition function
of this theory localizes on the solutions of least action, known as the BPS solutions, given
the topological properties of the fields. These solutions include the
instanton (or anti-self dual) solutions: $F=-*F$. Moreover, the
coefficients of the partition function can be shown to equal the Euler
numbers of instanton moduli spaces if the spaces
are smooth \cite{Vafa:1994tf}. If the moduli space is not smooth,
the definition of the suitable Euler number is rather intricate,
although it is clear physically that a well-defined integer should
exist. We simply refer to this quantity as the ``BPS invariant''.

The localization to BPS solutions allows to explicitly
compute the partition function using mathematical
techniques. Instanton moduli spaces are more abstractly described as
moduli spaces of semi-stable coherent sheaves with respect to the
K\"ahler modulus (or polarization) $J$ of the surface, see for example
\cite{Huybrechts:1996}. The Chern character of the sheaf $E$ is determined in terms of
the $U(r)$ field strength $F$ by: 
$$\mathrm{ch}(E)=r+\frac{i}{2\pi}
\mathrm{Tr}\,F+\frac{1}{8\pi^2}\mathrm{Tr}\,F\wedge F.$$
We abbreviate the topological classes of the sheaf by
$\Gamma=(r,\mathrm{ch}_1,\mathrm{ch}_2)$. Another often used quantity
in the context of sheaves is the discriminant $\Delta(E)$ defined by $\frac{1}{r(E)}(c_2(E)-\frac{r(E)-1}{2r(E)}c_1(E)^2)$. We
refer to the references for more details and definitions
\cite{Manschot:2011dj, Manschot:2011ym, Manschot:2010nc}. 

In this mathematical context it is actually possible to compute
instead of the Euler number the more refined Poincar\'e polynomial,
enumerating the Betti numbers of the moduli spaces. This quantity can
probably be obtained from the $\cN=4$ Yang-Mills path integral by
introducing a potential for an $R$-charge of the theory. We continue
by defining the refined BPS invariants $\Omega(\Gamma,w;J)$ in
an (informal) way, let
$p(\cM_J(\Gamma),w)=\sum_{i=0}^{2\dim_\mathbb{C}(\cM_J(\Gamma))}b_i(\cM_J(\Gamma))\,w^i$, be the Poincar\'e
polynomial of the moduli space of semi-stable sheaves for polarization
$J$. Then we define
\be
\label{eq:refw}
\Omega(\Gamma,w;J):=\frac{w^{-\dim_\mathbb{C}\cM_J(\Gamma)}}{w-w^{-1}}\,
p(\cM_J(\Gamma),w). 
\ee
The numerical invariant $\Omega(\Gamma;J)$ is defined by the limit $w\to -1 $:
\be
\Omega(\Gamma;J)=\lim_{w\to -1 }(w-w^{-1})\, \Omega(\Gamma,w;J)=(-1)^{\dim_\mathbb{C}\cM_J(\Gamma)}\chi(\cM_J(\Gamma)).
\ee
The rational refined invariants are defined by \cite{Manschot:2010nc}:
\be
\label{eq:rational}
\bar \Omega(\Gamma,w;J)=\sum_{m|\Gamma}
\frac{\Omega(\Gamma/m,-(-w)^m;J)}{m}.  
\ee

The generating function of refined BPS invariants for rank $r$ sheaves on the
surface $P$ takes the following form \cite{Vafa:1994tf, Minahan:1998vr,
  Manschot:2011dj, deBoer:2006vg}:
\begin{eqnarray}
\label{eq:genfunction}
\mathcal{Z}_r(y,z,\tau;P,J)&=&\sum_{c_1, c_2}\,\bar \Omega(\Gamma,w;J)\,(-1)^{rc_1\cdot K_P}\,\non\\
&&\times q^{r \Delta(\Gamma) -\frac{r\chi(P)}{24}-\frac{1}{2r}(c_1+rK_P/2)^2_-} \bar q^{
  \frac{1}{2r}(c_1+rK_P/2)^2_+} e^{-2\pi i \bar y \cdot (c_1+rK_P/2)},\non
\end{eqnarray}
where $y \in H_2(P,\mathbb{C})$, $w=\exp(2\pi i z)$ and $z\in
\mathbb{C}$. Furthermore, $K_P$ is the canonical class, and $\chi(P)$ the Euler
number of the surface $P$. Due to the decomposition $U(r)=U(1)\times
SU(r)$, this generating function can be decomposed in a set of theta
functions $\Theta_{r,\mu}(y,\tau;P)$, summing over $U(1)$-fluxes,
and functions $h_{r,\mu}(z,\tau;P,J)$ capturing the $SU(r)$ sector of
the gauge theory with magnetic flux given by
$\mu$:
\be
\label{eq:thetadecomp}
\mathcal{Z}_r(y,z,\tau; P, J)=\sum_{\mu\in H^2(P,\mathbb{Z}/r \mathbb{Z})} h_{r,\mu}(z,\tau;P,J)\,\overline{\Theta_{r,\mu}(y,\tau;P)},
\ee
with $\Theta_{r,\mu}(y,\tau;P)$ given by:
\be
\label{eq:ThetaS}
\Theta_{r,\mu}(y,\tau;P)=\sum_{\bfk \in H^2(P,r \mathbb{Z})+rK_P/2+\mu} (-1)^{r\bfk
  \cdot K_P} q^{\bfk^2_+/2r}\bar q^{-\bfk^2_-/2r} e^{2\pi iy\cdot
  \bfk },
\ee
 and:
\be
h_{r,\mu}(z,\tau;P,J)=\sum_{c_2} \bar \Omega(\Gamma,w;J)\,q^{r\Delta(\Gamma)-\frac{r\chi(P)}{24}}.
\ee
Mathematically, this decomposition is the consequence of the
isomorphism between moduli spaces of sheaves induced by tensoring the
sheaf by a line bundle. In the two-dimensional theory, this decomposition is understood as a
spectral flow symmetry. 

The partition function of the (numerical) BPS invariants is obtained
as the limit $\mathcal{Z}_r(y,\tau; P, J)=\lim_{w\to
  -1}(w-w^{-1})\,\mathcal{Z}_r(y,z,\tau; P, J)$. Physical
arguments such as $S$-duality suggest that $\mathcal{Z}_r(y,\tau; P, J)$ transforms as
a multi-variable Jacobi form of weight $(-\frac{3}{2},\frac{1}{2})$
under $SL(2,\IZ)$. Since $\Theta_{r,\mu}(y,\tau;P)$ is a vector-valued Jacobi
form of weight $(\half,\half (b_2(P)-1))$, the weight of
$h_{r,\mu}(\tau;P,J)$ is $-\chi(P)/2$.

An intriguing relation is the blow-up formula which relates for a
surface $P$ and its blow-up $\tilde P$ the generating functions 
$h_{r,\mu}(z,\tau;P,J)$ and $h_{r,\mu}(z,\tau;\tilde P,J)$\cite{Yoshioka:1994,
  Manschot:2011ym}. Let $\phi:\tilde P\to P$ denote the map between
the two surfaces and $\phi^*$ be the pull-back acting on differential
forms. Let moreover $\bc$ be the additional two-cycle which arises
from the blow-up (the exceptional divisor of $\phi$). Then the blow-up
formula gives an expression for $h_{r,\phi^*c_1-k\bc}(z,\tau;\tilde P,J)$ in terms of $h_{r,c_1}(z,\tau;P,J)$.
We will explain this relation for the greatest common divisor
$\gcd(r,c_1)=1$. 
In this case:
\be 
\label{eq:blowup}
h_{r,\phi^*c_1-k\bc}(z,\tau;\tilde P,J)=B_{r,k}(z,\tau)\,h_{r,c_1}(z,\tau;P,J)\ ,
\ee 
with 
\be
B_{r,k}(z,\tau)=\frac{1}{\eta(\tau)^r} \sum_{\sum_{i=1}^r a_i=0 \atop a_i\in \mathbb{Z}+\frac{k}{r}} q^{-\sum_{i<j}a_ia_j}\,w^{\sum_{i<j}a_i-a_j}.
\ee
For  $\gcd(r,c_1)>1$ the manipulations are more
intricate due to strictly semi-stable sheaves \cite{Yoshioka:1994,
  Manschot:2011ym}.

\section{Formulation and tests of the conjecture}

We start by explicitly formulating the conjecture. Let $\cM_r=\mathbb{R}^3 \times (S^1\times \widetilde \cM_r)/ \mathbb{Z}^r$ be the $r-$monopole moduli space. For $r=2$, the relative moduli space ${\widetilde \cM}_{r=2}$ is the
Atiyah-Hitchin manifold. We claim that the
elliptic genus of the $(0,4)$ sigma model with target space $\cM_r$ is given by partition function of topologically twisted
$\cN=4$ Yang-Mills on $\IB_0\cong \bP^1\times \bP^1$. This statement in fact also appeared in \cite{arXiv:1107.2847}. Here, we elaborate on it and extend it to include vector bundles over $\cM_r$, or, in five dimensional language, flavor hypermultiplets.

Five dimensional gauge theory with $N_f$ hypermultiplets in the fundamental representation gives us 
naturally an $SO(2rN_f)$ bundle over $\cM_r$. The corresponding sigma model contains
$rN_f$ chiral bosons (or $2rN_f$ chiral fermions) on the left moving side which are sections
of an $SO(2rN_f)$ bundle, as explained in Section 2. There is a global $SO(2N_f)$ flavor symmetry group, and the structure is schematically displayed in (\ref{eq:sigmatarget}): 
\be
\label{eq:sigmatarget}
\xymatrixcolsep{-.5pc}
\xymatrix{& O(r) \ar[d] &\hspace{-.7cm} \times SO(2N_f)\\ \mathbb{R}^3 \times &
  (S^1\times \widetilde \cM_r)/ \mathbb{Z}^r}.
\ee

\noi {\bf Conjecture}\newline
The elliptic genus of the  $(0,4)$ sigma model based on 
(\ref{eq:sigmatarget}) is given by the partition function of $\cN=4$
Yang-Mills  on $\IB_{N_f}$. 
\newline
 
Clearly, the conjecture simplifies a lot for $r=1$. In that case, the
partition function can be explicitly computed on both
sides, providing a non-trivial test of the $SO(2N_f)$ information of the
2d theory. We will also discuss the conjecture for $r=2$, and the
relation with four-dimensional $\cN=2$ gauge theory.  

\subsection{Derivation for $r=1$}
\label{subsec:derivr=1}

Here we will present a proof of the conjecture for $r=1$. We will start by defining the $(0,4)$ elliptic genus and computing it for the case of the magnetic string sigma model with magnetic monopole charge $1$, with action given by \eqref{freeCFT}. As a next step we will proceed to determine the $\cN=4$ SYM partition function on $\IB_{N_f}$ and compare the two results.

\subsubsection*{(0,4) CFT} 
The elliptic genus of the $(0,4)$ SCFT is defined by \cite{Minahan:1998vr, Gaiotto:2006wm, deBoer:2006vg, Denef:2007vg}:
\be 
\label{eq:ellgenus}
\cZ_{\mathrm{CFT}}(\tau,y)=\Tr_{\textrm{R}}\, \left[\half F^2(-1)^F q^{L_0-\frac{c_L}{24}} \bar q^{\bar L_0-\frac{c_R}{24}} e^{2\pi i y_i J_0^i}\right]\ .
\ee
Here, $F$ is the fermion number of the adjoint fermions and is inserted to soak up the 4 fermionic zero modes of the monopole background.
The $J_0^i\,, \,  i = 1 , \cdots, N_f$ are zero modes of currents corresponding to the Cartan subalgebra of $SO(2N_f)$ and the $y_i$ correspond to Wilson lines as discussed in Section \ref{sec:masshyp}. As usual, we have that $q=\exp{2\pi i \tau}$.

In this section we want to compute the index for charge $r=1$ monopoles in the background of $N_f$ massless flavors.
The index receives contributions from the 3 non-compact bosons whose partition function is given by
\begin{equation}
	\cZ_{\IR^3}=\Tr\, \left[q^{L_0-\frac{3}{24}}\bar q^{\bar L_0-\frac{3}{24}}\right]=\frac{1}{\tau_2^{3/2}}\frac{1}{|\eta(\tau)|^6} .
\end{equation}
Furthermore, there is the contribution from the compact scalar
parametrized by $\theta$ and the $2N_f$ left-moving fermions $\chi^A,
~ A=1,\cdots, 2N_f$, which together form the so called M\"obius bundle
(in the following abbreviated by "MB"). Its partition function is
given by 
\begin{equation}
	\cZ_{\textrm{MB}} = \Tr\, \left[\bar{q}^{\bar{L}_0-\frac{1}{24}} q^{L_0-\frac{1+N_f}{24}} e^{2\pi i y_i J_0^i}\right]=\frac{\Theta_{\textrm{MB}}(\tau,y)}{\overline{\eta(\tau)} \eta(\tau)^{N_f+1}}\ .
\end{equation}
A last factor comes from the 4 right-moving adjoint fermions and is of the form
\begin{equation}
	\Tr\, \left[(-1)^{F} \bar{q}^{L_0-\frac{2}{24}} e^{2\pi i z F}\right]=\frac{\overline{\vartheta_1(\tau,z)}^2}{\overline{\eta(\tau)}^2}\ .
\end{equation}

The elliptic genus (\ref{eq:ellgenus}) is now computed by multiplying
the three contributions, taking twice the derivative to $z$ at
$z=0$. One obtains:
\be
\cZ_{\mathrm{CFT}}(\tau)=\frac{\Theta_{\textrm{MB}}(\tau)}{\eta(\tau)^{N_f+4}}\ .
\ee

The main difficulty lies in the computation of $\cZ_{\textrm{MB}}$. As explained in Section 2.1, the constraint \eqref{cftconstr} leads to an orbifolded CFT, where we mod out by the group $G=\{1,g\}$ generated by the identity and $g=(-)^{n_e + F}$. Denoting the moduli space parametrized by the collective coordinates $\theta$ and $\chi$ by $\cM$ we have to compute 
\begin{displaymath}
	\cZ_{\textrm{MB}} = \cZ_{\cM/G}  = \frac{1}{|G|} (\cZ_{\cM}[^1_1] + \cZ_{\cM}[^g_1] + \cZ_{\cM}[^1_g] + \cZ_{\cM}[^g_g])\ ,
\end{displaymath}
where $\cZ_{\cM}[^x_y]$ corresponds to the partition function of the magnetic string twisted by $x \in G$ in the time direction and by $y \in G$ in the space direction. Thus we obtain
\begin{equation}
	\cZ_{\cM/G} = \frac{1}{2} \textrm{Tr}_1\left[(1 + (-)^{n_e + F}) q^{L_0 - \frac{c_L}{24}} \bar{q}^{\bar{L}_0 - \frac{c_R}{24}}\right] + \frac{1}{2} \textrm{Tr}_g\left[(1+(-)^{n_e + F}) q^{L_0 - \frac{c_L}{24}} \bar{q}^{\bar{L}_0 - \frac{c_R}{24}}\right]\ ,
\end{equation}
where $\textrm{Tr}_1$ denotes the trace in the untwisted sector and $\textrm{Tr}_g$ is the trace in the $g$-twisted sector. Denoting by $P_c$ the projection operator to states satisfying the condition $c$ the trace in the untwisted sector becomes 
\begin{eqnarray}
	\frac{1}{2} \textrm{Tr}_1\left[\cdots\right] 
	& = & \textrm{Tr}_1\left[\langle \theta'|\otimes \langle \chi'| P_{\textrm{even}~n_e} P_{\textrm{even}~F} ~q^{L_0 - \frac{c_L}{24}} \bar{q}^{\bar{L}_0 - \frac{c_R}{24}}|\theta\rangle \otimes |\chi\rangle\right] \nonumber \\
	& ~ & + \textrm{Tr}_1\left[\langle \theta'|\otimes \langle \chi'| P_{\textrm{odd}~n_e} P_{\textrm{odd}~F}  ~q^{L_0 - \frac{c_L}{24}} \bar{q}^{\bar{L}_0 - \frac{c_R}{24}}|\theta\rangle \otimes |\chi\rangle\right] \nonumber ,
\end{eqnarray}
which can be rewritten as
\begin{eqnarray}
	~ & ~ & \textrm{Tr}_1 \left[\langle \theta'|\frac{1}{2}(1+(-)^{n_e})q^{L_0 - \frac{c_L}{24}} \bar{q}^{\bar{L}_0 - \frac{c_R}{24}}|\theta\rangle\right] \cdot \textrm{Tr}_1 \left[\langle \chi'| \frac{1}{2}(1 + (-)^{F})q^{L_0 - \frac{c_L}{24}} |\chi\rangle \right] \nonumber \\
	~ & + & \textrm{Tr}_1 \left[\langle \theta'|\frac{1}{2}(1-(-)^{n_e})q^{L_0 - \frac{c_L}{24}} \bar{q}^{\bar{L}_0 - \frac{c_R}{24}}|\theta\rangle\right] \cdot \textrm{Tr}_1 \left[\langle \chi'| \frac{1}{2}(1 - (-)^{F})q^{L_0 - \frac{c_L}{24}} |\chi\rangle \right]. \nonumber
\end{eqnarray}
The expression for the twisted sector is completely analogous and is obtained by replacing $\textrm{Tr}_1$ by $\textrm{Tr}_g$ in the above formula. The untwisted sector corresponds to periodic boundary conditions and can therefore be identified with the Ramond sector. We can now evaluate
\begin{eqnarray}
	\textrm{Tr}_R\left[\langle \chi'| \frac{1}{2}(1 + (-)^F) q^{L_0 - \frac{c_L}{24}} e^{2\pi i y_i J_0^i}|\chi\rangle\right] 
	& = & \frac{1}{2} \left[\prod_{i=1}^{N_f} \frac{\vartheta_2(\tau, y_i)}{\eta(\tau)} + \prod_{i=1}^{N_f} \frac{\vartheta_1(\tau,y_i)}{\eta(\tau)}\right], \nonumber \\
    \textrm{Tr}_R\left[\langle \chi'| \frac{1}{2}(1 - (-)^F)  q^{L_0 - \frac{c_L}{24}} e^{2\pi i y_i J_0^i}  |\chi\rangle\right] 
	& = & \frac{1}{2} \left[\prod_{i=1}^{N_f} \frac{\vartheta_2(\tau, y_i)}{\eta(\tau)} - \prod_{i=1}^{N_f} \frac{\vartheta_1(\tau,y_i)}{\eta(\tau)}\right]. \nonumber \\
\end{eqnarray}
The expressions for the twisted sector are obtained by noting that the mode expansion of the $\chi^A$ becomes shifted by half-integers due to 
\begin{equation}
	\chi^A(x + 2\pi, t) = - \chi^A(x,t), \quad A=1, \cdots, 2N_f.
\end{equation}
Therefore, the trace has to be computed in the NS sector:
\begin{eqnarray}
\label{eq:thetachi}
	\textrm{Tr}_{NS}\left[\langle \chi'| \frac{1}{2}(1 + (-)^F) q^{L_0 - \frac{c_L}{24}} e^{2\pi i y_i J_0^i} |\chi\rangle\right] 
	& = & \frac{1}{2} \left[\prod_{i=1}^{N_f} \frac{\vartheta_3(\tau, y_i)}{\eta(\tau)} + \prod_{i=1}^{N_f} \frac{\vartheta_4(\tau,y_i)}{\eta(\tau)}\right], \nonumber \\
	\textrm{Tr}_{NS}\left[\langle \chi'| \frac{1}{2}(1 - (-)^F)  q^{L_0 - \frac{c_L}{24}} e^{2\pi i y_i J_0^i} |\chi\rangle\right] 
	& = & \frac{1}{2} \left[\prod_{i=1}^{N_f} \frac{\vartheta_3(\tau, y_i)}{\eta(\tau)} - \prod_{i=1}^{N_f} \frac{\vartheta_4(\tau,y_i)}{\eta(\tau)}\right]. \nonumber \\
\end{eqnarray}
Next, we want to compute the trace of the compact boson sector. In order to proceed we recall here some facts about translation orbifold blocks for the compact boson CFT (see for example Appendix B of \cite{Kiritsis:1997hj}). This CFT is a 1-dimensional toroidal CFT with symmetry group $O(1,1)$. The transformations associated with it are arbitrary lattice translations which act on a state with momentum $p$ and winding number $w$ as
\begin{equation}
	g_{\textrm{translations}} = \textrm{exp}\left[2\pi i (p \phi_1 + w \phi_2) \right],
\end{equation}
where $\phi_1$ and $\phi_2$ are rational numbers. This results in a freely acting discrete group of finite order $N$. When modding out by this symmetry the resulting orbifold has twisted sectors with shifted momentum and winding modes of the form
\begin{equation}
	p \mapsto p + n \phi_2, \quad w \mapsto w + n \phi_1, \quad n = 0, \cdots,~N-1.
\end{equation}
In our case we are modding out by the symmetry 
\begin{equation}
	g = \exp\left[2 \pi i \frac{n_e}{2}\right] = \exp\left[2 \pi i \left(\frac{p}{2} - \frac{8-N_f}{4} w\right)\right],
\end{equation}
where use has been made of the identities (\ref{wpbasis}). This results in the following shifts of $p$ and $w$ in the twisted sector:
\begin{equation} \label{twshifts}
	p \mapsto p + \frac{8-N_f}{4}, \quad w \mapsto w + \frac{1}{2}.
\end{equation}
Now we are in the position to write down the full partition function. It will consist of four pieces, the untwisted sector with even $n_e$ and odd $n_e$, as well as the twisted sector with even $n_e$ and odd $n_e$. The difference between the twisted and untwisted sector will be the shift (\ref{twshifts}) in momentum and winding number. The full result then looks as follows
\begin{eqnarray} \label{eq:2dresult1}
	\cZ_{\cM/G} & = & \cZ_{S^1}^{\textrm{untw},~n_e~\textrm{even}} \cdot \frac{1}{2}\left[\left(\frac{\vartheta_2}{\eta}\right)^{N_f} + \left(\frac{\vartheta_1}{\eta}\right)^{N_f}\right]  \nonumber \\
	~                  & ~ & + ~\cZ_{S^1}^{\textrm{untw},~n_e~\textrm{odd}} \cdot \frac{1}{2}\left[\left(\frac{\vartheta_2}{\eta}\right)^{N_f} - \left(\frac{\vartheta_1}{\eta}\right)^{N_f}\right]  \nonumber \\
	~                  & ~ & + ~\cZ_{S^1}^{\textrm{tw},~n_e~\textrm{even}} \cdot \frac{1}{2}\left[\left(\frac{\vartheta_3}{\eta}\right)^{N_f} + \left(\frac{\vartheta_4}{\eta}\right)^{N_f}\right] \nonumber \\
	~                  & ~ & + ~\cZ_{S^1}^{\textrm{tw},~n_e~\textrm{odd}} \cdot \frac{1}{2}\left[\left(\frac{\vartheta_3}{\eta}\right)^{N_f} - \left(\frac{\vartheta_4}{\eta}\right)^{N_f}\right] 
\end{eqnarray}
The above can be written more conveniently in terms of $\cZ_{S^1}^{a,b}$ where $a, b \in \{0,1\}$. In this notation $a=0$ stands for the untwisted sector and $a=1$ for the twisted sector, $b=0$ corresponds to even $n_e$ and $b=1$ to odd $n_e$. Then we have
\begin{equation} \label{eq:2dresult2}
	\cZ_{S^1}^{a,b} = \frac{1}{|\eta (\tau)|^2} \left[\sum_{\stackrel{w \in \IZ + \frac{a}{2}~,}{p \in \frac{(8-N_f)}{2} w + 2\IZ + b + \frac{1}{4}(1-(-)^{N_f})(1-a)}} q^{\frac{1}{2} p_L^2} \bar{q}^{\frac{1}{2} p_R^2}\right].
\end{equation}
The shift $\frac{1}{4}(1-(-)^{N_f})(1-a)$ is included to make sure that for odd $N_f$ the constraint $2pw \in 2\IZ$ is satisfied in the untwisted sector.

\subsubsection*{$\cN=4$ Yang-Mills}

We will now determine the partition function $\mathcal{Z}_r(y,z,\tau;\IB_{n},J)$ (\ref{eq:genfunction}) for $r=1$. 
These simplify considerably: the sum over $\Lambda/r\Lambda$ reduces to a
single term, and $h_{1,c_1}(z,\tau;\IB_{n})$ does not depend on $J$
since all rank 1 sheaves are stable. The moduli spaces of these
sheaves correspond to the Hilbert scheme of points on $\IB_{n}$. Therefore,
$h_{1,c_1}(z,\tau;\IB_{n})$ is given by \cite{Gottsche:1990}: 
\be
h_{1,c_1}(z,\tau;\IB_{n})=\frac{i}{\theta_1(\tau,2z)\,\eta(\tau)^{b_2(\IB_{n})-1}}.\non 
\ee
This gives for the numerical invariants:
\be
h_{1,c_1}(\tau;\IB_{n}))=\frac{1}{\eta(\tau)^{\chi(\IB_{n})}}=\frac{1}{\eta(\tau)^{4+n}}\ ,
\ee
since $\chi(\IB_{n})=2+b_2(\IB_{n})=4+n$.

Next, we discuss in more detail the other functions in
(\ref{eq:thetadecomp}), namely the theta functions
$\Theta_{r,\mu}(y,\tau;\IB_n)$ summing over the lattice $\Lambda$. 
In order to relate these functions to the two-dimensional results, we consider $\Lambda$ as the gluing of the
lattices $A$ and $D$ as described in Section \ref{subsec:surfacelattice}. In order to compare with the two-dimensional result we choose the parametrization (\ref{eq:Jfieldtheory}) for $J$ and set the masses $m_i$ to zero. Then the $\Theta_{r,\mu}(y,\tau;\IB_n)$
factorizes as the restriction of $J$ to $D$ vanishes; it takes the form: 
\be
\label{eq:thetafactor}
\Theta_{r,\mu}(y,\tau;\IB_n)=\sum_{i=0}^3 \Theta_{rA,\bg_i+\mu}(y,\tau)\,\overline{\Theta_{rD_n,\bg_i+\mu}(y,\tau)},
\ee
where $\bg_i$ are the gluing vectors. Moreover, $\Theta_{rA,\mu}(y,\tau)$ is given by:
\be\label{theta-A-lattice}
\Theta_{rA,\mu}(y,\tau)=\sum_{\bfk\in \frac{1}{2}\bfa_1+\mu +(r\mathbb{Z})^2} (-1)^{r\bfa_1\cdot\bfk} q^{\frac{\bfk^2_+}{2r}}\bar q^{-\frac{\bfk^2_-}{2r}}e^{2\pi i y\cdot \bfk}
\ee
where the quadratic form $\bfk^2$ is obtained from the $A$ matrix
(\ref{eq:Amatrix}). Similarly, $\Theta_{rD_n,\mu}(y,\tau)$ is given by
\be
\Theta_{rD_n,\mu}(y,\tau)=\sum_{\bfk\in \mu+(r\mathbb{Z})^{n}} q^{\frac{\bfk^2}{2r}}\,e^{2\pi i y\cdot \bfk},
\ee
and the quadratic form $\bfk^2$ is here obtained from the $D_n$ Cartan matrix. 
We have left implicit in the formulas above that $\bg_i+\mu$ and
$y$ should be restricted to the lattice $A$ and $D$ for $\Theta_{rA,\bg_i+\mu}(y,\tau)$ and
$\overline{\Theta_{rD_n,\bg_i+\mu}(y,\tau)}$ respectively.

For $r=1$ and $y=0$, one finds for the $D_n$ theta functions the following:
\begin{eqnarray}
\label{eq:Dnthetas}
\Theta_{D_n,\bg_0}(\tau)&=&\frac{1}{2}\left( \vartheta_3^n+\vartheta_4^n\right)=1+2n(n-1)\,q+\dots,\\
\Theta_{D_n,\bg_1}(\tau)&=& \frac{1}{2} \vartheta_2^n= 2^{n-1} q^{\frac{n}{8}}+\dots,\\
\Theta_{D_n,\bg_2}(\tau)&=&\half \left( \vartheta_3^n-\vartheta_4^n\right)=2n\, q^\half+\dots,\\ 
\Theta_{D_n,\bg_3}(\tau)&=&\frac{1}{2} \vartheta_2^n=2^{n-1} q^{\frac{n}{8}}+\dots,
\end{eqnarray}
i.e. precisely the theta functions of the current algebra $\widehat{so}(2n)_1$. The
first coefficients in the $q$-expansion are the dimensions of $SO(2n)$
representations. Comparing with (\ref{eq:thetachi}), we observe that
these $D_n$ theta functions precisely correspond to  the ones obtained
from the various orbifold sectors in two dimensions!

It remains, for the identification of the four-dimensional with the
two-dimensional partition function, to identify the $\Theta_{A,\bg_i}$
with the different twisted and untwisted theta-functions of
$\cZ_{S^1}$. In order to perform this comparison we need to rewrite
the summation in $\Theta_{A,\bg_i}$ in the same units as the summation
involved in $\cZ_{S^1}$. This means identifying the vector $\mathbf{k}
\in H^2(\IB_n,\IZ) + \frac{r K_{\IB_n}}{2}$ with the charge vector of
the five-dimensional field theory as in (\ref{eq:chargevector}).

In this subsection, we will work in the massless case, $m_i=0$ and for
magnetic charge $r=1$. One then easily computes with (\ref{eq:Jfieldtheory})
\begin{equation}
\frac{1}{2} J^2=\frac{T}{\sqrt 2}\ ,\qquad J\cdot \bfk=n_IZ_I+n_e\phi\ ,\qquad -\bfk \cdot K_{\IB_{N_f}}=n_e+(8-N_f)n_I\ ,
\end{equation}
so we recover in the second equation the central charge of the dyonic instanton with instanton charge $n_I$ and electric charge $n_e$. It now follows straightforwardly that 
\begin{equation}\label{k+2}
\bfk_+^2=\frac{(n_e\phi+n_IZ_I)^2}{\sqrt{2} T}\ ,
\qquad 
\end{equation}
and furthermore, we have the important identity
\begin{equation}\label{ksquared}
\bfk_{|_A}^2=\bfk_+^2-(-\bfk_{|_A,-}^2)=2n_I\Big(n_e+\frac{(8-N_f)}{2}\,n_I\Big)\ ,
\end{equation}
where we have restricted the charge vector $\bfk$ to the lattice $\Lambda_A$, which is needed in the calculation of the partition function. Comparing the partition function on the $A$-lattice \eqref{theta-A-lattice} with the partition function for a conformal field theory, we identify left- and right moving momenta as
\begin{equation} \label{eq:windmomid}
p_R^2=\bfk_+^2\ ,\qquad p_L^2=-\bfk_{|_A,-}^2\ ,
\end{equation}
with the identification of $p_R$ and $p_L$ as in (\ref{eq:pRL}).

To compute the partition function more explicitly, we have to find the theta-functions corresponding to the different gluing vectors $\bg_i$ restricted to the $A$-lattice and identify them with twisted and untwisted sectors. For the vector $\bfk$ restricted to the $A$-lattice $\bfk_A$, this implies
that the coefficients are rational: 
\begin{equation}
\bfk_A \in (\textstyle \half,0) + {\bg}_{i,A}+A\ ,
\end{equation}
where ${\bg}_i; i=0,...,3$ is one of the four gluing
vectors. Comparing with (\ref{eq:chargevector}) gives for the charges
\begin{equation}\label{shifts-nenI}
n_I\in \frac{1}{2}+{\bg}_{i,I} + \mathbb{Z}\ ,\qquad n_e\in 2 {\bg}_{i,e}+  2\mathbb{Z}\ ,
\end{equation}
where ${\bg}_{i,I}$ is the $I$-component of the gluing vector ${\bf
  g}_i$, etc. Accordingly, Eq. (\ref{eq:windmomid}) implies that
also the winding and momentum modes are rational:
\begin{equation}
w\in \frac{1}{2}+{\bg}_{i,I}+\mathbb{Z}\ ,\qquad   p\in 2{\bg}_{i,e} + \frac{(8-N_f)}{2}\,\Big(\frac{1}{2}+{\bg}_{i,I}\Big)+\mathbb{Z}\ .
\end{equation}
We now look in more detail at the properties of the gluing vectors
that are given in Section \ref{subsec:surfacelattice}. We split them into two sectors, the
twisted sector, defined by ${\bg}_0$ and ${\bg}_2$, and the
untwisted sector, spanned by ${\bg}_1$ and ${\bg}_3$. These
sectors will correspond to the twisted and untwisted sectors of the
orbifold CFT to which we compare at the end of this subsection.  

In the twisted sector, we see that ${\bg}_{i,I}$ is always zero, so we
get half-integer shifts of the winding modes. Furthermore, in this
sector $2{\bg}_{0,e}=0$ and $2{\bg}_{2,e}=1$. Therefore, the
twisted sector corresponds to $n_I \in \IZ + \frac{1}{2}$ and further
splits into two sub-sectors with even and odd $n_e$. So the shifts we
obtain for $w$ and $p$ are 
\begin{equation}
twisted\,\,\, sector: \qquad w\in \frac{1}{2}+\mathbb{Z}\ ,\qquad p \in \pm \frac{N_f}{4} +\mathbb{Z}\ .
\end{equation}
Comparing to the 2d computation we see that this exactly matches the result (\ref{eq:2dresult2})!

In the untwisted sector, we have ${\bg}_{i,I}=1/2$ such that $n_I\in \mathbb{Z}$. For the electric component, one can check
explicitly that for $N_f=0,1,2,3$, the momentum mode gets shifted by
an amount that can be absorbed in either an even shift in $n_e$ or an
integer shift in $n_I$. Hence effectively,   we can drop these shifts
in the lattice sum. We conclude that in the untwisted sector 
\begin{equation}
untwisted\,\,\, sector: \qquad w \in \mathbb{Z} \ ,\qquad p\in
\mathbb{Z}, \quad \mathrm{or}\quad p\in \frac{N_f}{2} +\mathbb{Z}\ .
\end{equation}
Again we see that we obtain the same mode expansion as encoded in (\ref{eq:2dresult2}). Also, one observes that the difference between ${\bg}_{1,e}$ and ${\bg}_{3,e}$ just results in a shift of $n_e$ by $1$, leading thus to the two subsectors with odd and even $n_e$. We therefore see that the two partition functions exactly match and thereby confirm our conjecture for $r=1$.\footnote{As the 2d partition function does not contain a parameter $y$, we set $y = -\frac{1}{2} \mathbf{a}_1$ to cancel the factor $(-1)^{\mathbf{a_1} \cdot \mathbf{k}}$ in the definition of $\Theta_{A}(y,\tau)$.}

\subsection{Rank $r\geq 2$}
\label{sec:rgeq2}
This section discusses the proposed correspondence for higher rank
$r\geq 2$. On the $\mathcal{N}=4$ Yang-Mills side, the partition
functions for $r\geq 2$ can be determined explicitly. Unfortunately,
this is not feasible on the two-dimensional side. Therefore, the
partition function of $\mathcal{N}=4$ Yang-Mills provides a prediction for
the elliptic genus of the $(0,4)$ sigma model.

The partition functions for $r\geq 2$, can be determined using the techniques developed in \cite{Yoshioka:1994, Yoshioka:1995,
  Manschot:2010nc, Manschot:2011ym}. The key property of the
computation is that  $\IF_0$ is the product $\IP^1\times \IP^1$
(and more generally for Hirzebruch surfaces $\IF_n$ it is a
fibration). The BPS invariants can be determined if one chooses the
K\"ahler modulus $J$ such that one $\bP^1$ (the fibre $\bff$) is
infinitesimally small compared to the  other $\bP^1$ (the base $\bC$).
This choice of $J$ is called a suitable polarization. Using the jump
of the invariants across walls of marginal stability \cite{Yoshioka:1994, 0811.2435, MR2357325}, one can consequently compute the
invariants for other choices of $J$. The BPS-invariants of $\IB_{n}$
are determined from those of $\IF_0$ using the blow-up formula (\ref{eq:blowup}). 
 
We start with the computation of the partition functions for $\IB_0$. 
The polarization $J$ is parametrized by $J_{m_1,m_2}=m_1\bC+m_2\bff$, such that the suitable
polarization is given by $J_{\varepsilon,1}$ with $0<\varepsilon\ll 1$.  
For this choice the BPS-invariants vanish for all sheaves with $\gcd(r,
\bff\cdot c_1)=1$. If this condition is not satisfied, the BPS-invariants do not vanish and their computation is more involved due to
the presence of strictly semi-stable sheaves \cite{Yoshioka:1995,
  Manschot:2011ym}.  One finds for $h_{2,c_1}(z,\tau;J_{\varepsilon,1},\IB_0)$:
\begin{eqnarray}
&& \beta=1\mod 2: \non \\
&&\quad h_{2,\beta \bC-\alpha \bff}(z,\tau;J_{\varepsilon,1},\IB_0)=0,\qquad \\
&&(\alpha,\beta)=(1,0)\mod 2: \non \\
&&\quad h_{2,\beta \bC-\alpha \bff}(z,\tau;J_{\varepsilon,1}, \IB_0)=\frac{-i\,\eta(\tau)}{\vartheta_1(\tau,2z)^2\vartheta_1(\tau,4z)}+\frac{w^2}{1-w^4}h_{1,0}(z,\tau;\IB_0)^2,\non \\
&&(\alpha,\beta)=(0,0)\mod 2:\non \\
&&\quad h_{2,\beta \bC-\alpha \bff}(z,\tau;J_{\varepsilon,1},\IB_0)=\frac{-i\,\eta(\tau)}{\vartheta_1(\tau,2z)^2\vartheta_1(\tau,4z)}+\left(\frac{1}{1-w^4}-\frac{1}{2}\right)h_{1,0}(z,\tau;\IB_0)^2.\non
\end{eqnarray}
Note that the coefficients of $h_{2,{\bf 0}}(z,\tau;J_{\varepsilon,1})$
are rational due to the multi-covering formula (\ref{eq:rational}).
The modular properties of this generating function appear to be more
elegant than the ones for integer invariants. However, why these rational coefficients appear in the
generating function from the point of view of the $\cN=4$ Yang-Mills
theory or the two-dimensional field theory is not well understood. 

The partition function for more general $J_{m_1,m_2}$ is given by:
\begin{eqnarray}
\label{eq:hJ}
h_{2,\beta \bC-\alpha \bff}(z,\tau;J_{m_1,m_2},\IB_0)&=&h_{2,\beta \bC-\alpha
  \bff}(z,\tau;J_{\varepsilon,1};\IB_0)\\
&&+\frac{1}{4} h_{1,0}(z,\tau;\IB_0)^2\,\sum_{(a,b)=-(\alpha,\beta)\mod 2}
(\sgn(bm_2-am_1)-\sgn(b-a\varepsilon))\non \\
&&\times \,(w^{2b-2a}-w^{-2b+2a})\,q^{\frac{1}{2}ab}.\non
\end{eqnarray}
One can verify that for $J_{1,\varepsilon}$ this function vanishes for
$\alpha=1\mod 2$, which is expected due to the symmetry of $\bff$ and
$\bC$. Of special interest is the choice $(m_1,m_2)=(1,1)$ which
corresponds to $J=-K_{\IF_0}$. At this point, a geometric sum can be a carried out in
(\ref{eq:hJ}) giving: 
\begin{eqnarray}
&&h_{2,0}(z,\tau;J_{1,1}, \IB_0)=\frac{-i\eta(\tau)}{\vartheta_1(\tau,2z)^2\,\vartheta_1(\tau,4z)}\\
&&\qquad \qquad+ h_{1,0}(z,\tau;\IB_0)^2\,\left(\sum_{b\,\,\mathrm{even}}\frac{q^{\frac{1}{2}
      b^2}}{1-w^4q^b}-\frac{1}{2}q^{\frac{1}{2} b^2} \right)\non \\
&&h_{2,\beta \bC-\alpha \bff}(z,\tau;J_{1,1}, \IB_0)=h_{1,0}(z,\tau)^2\,\left(\sum_{b\,\,\mathrm{odd}}\frac{q^{\frac{1}{2}
      b^2+\frac{1}{2}b}w^2}{1-w^4q^b} \right), \quad \alpha+\beta=1\non \\
&&h_{2, \bC- \bff}(z,\tau;J_{1,1},\IB_0)= h_{1,0}(z,\tau;\IB_0)^2\,\left(\sum_{b\,\,\mathrm{odd}}\frac{q^{\frac{1}{2}
      b^2}}{1-w^4q^b}-\frac{1}{2}q^{\frac{1}{2} b^2} \right). \non 
\end{eqnarray}
The functions in brackets are (up to multiplication by a theta function) specializations of the Lerch-Appell sum \cite{Zwegers:2000}:
\be
\mu(u,v;\tau)=\frac{e^{i\pi u}}{\theta_1(v;\tau)}\sum_{n\in \mathbb{Z}} \frac{(-1)^n e^{\pi i(n^2+n)\tau+2\pi in v}}{1-e^{2\pi i n\tau+2\pi i u}}.
\ee

The surfaces $\IB_{n}$ are obtained by blowing up $n$ points of
$\IF_0$. Section \ref{subsec:surfacelattice} discussed how the lattice
$H_2(\IB_n,\mathbb{Z})$ is a gluing of a two-dimensional lattice $A$
and a $n$-dimensional lattice $D$. Therefore, the polarization $J$ can vary in more
directions. We consider here only variations of $J$ in the sublattice
$A$ and parametrize $J$ by $J_{m_1,m_2}=m_1\,\bfa_1+m_2\,\bfa_2\propto-\phi\,
K_{B_{n}}+\frac{1}{4g_5^2}\bff$, and the hypermultiplets are thus massless. As for $\IF_0$,
we start close to a boundary of the K\"ahler cone,
$J_{\varepsilon,1}$, where the BPS-invariants vanish if
$\gcd(r,c_1\cdot \bfa_2)>1$. They again do not vanish if this
condition is not satisfied, and can then be determined exactly. 
For illustration, we give two examples:
\begin{eqnarray}
h_{2,{\bf 0}}(z,\tau;J_{\varepsilon,1},\IB_n)&=& h_{2,c_1}(z,\tau;J_{0,1},\IB_n)\\
&&+\left(\frac{1}{1-w^4}-\frac{1}{2}
\right) h_{1,0}(z,\tau;\IB_n)^2\,\Theta_{2D_n,\boldsymbol{0}}(\tau)
\non \\
&& +\frac{w^2}{1-w^4}\,h_{1,0}(z,\tau;\IB_n)^2\,\Theta_{2D_n,\boldsymbol{1}}(\tau).\non
\end{eqnarray}
and for $c_1=\frac{n}{2} \bfa_2-\frac{1}{2}\sum_{i=1}^n \bd_i$:
\begin{eqnarray}
\label{eq:h23b}
h_{2,c_1}(z,\tau;J_{\varepsilon,1},\IB_n)&=& h_{2,c_1}(z,\tau;J_{0,1},\IB_n)+\frac{w}{1-w^2}\,h_{1,0}(z,\tau;\IB_3)^2\,\Theta_{2D_n,\frac{1}{2}\boldsymbol{1}}(\tau).
\end{eqnarray}
where
\be
h_{2,c_1}(z,\tau;J_{0,1},\IB_n)=\frac{-i\,\eta(\tau)}{\theta_1(\tau,2z)^2\,\theta_1(\tau,4z)}\prod_{i=1}^n B_{2,k_i}(z,\tau),\non
\ee
with $k_i=c_1\cdot \bc_i$. The product over $i=1,\dots,n$ on the right
hand side is due to the blow-up formula (\ref{eq:blowup}). Similarly to $n=0$, the point $J=-K_{B_{n}}$
is special for $n>0$,: the sum over walls between $J_{\varepsilon,1}$ and
$-K_{B_{n}}$ can be resummed to a specialization of a Lerch-Appel
function multiplied by a theta function. 

One can understand for general $r$ the blow-up formula from the two-dimensional perspective
from the $so(2rN_f)$ current algebra arising from the $SO(2rN_f)$
bundle over the monopole moduli space. The corresponding theta function
sums over an $rN_f$-dimensional lattice. The sum over an
$N_f$-dimensional sublattice gives $\Theta_{rD_n,\mu}(y,\tau)$, and
the remaining $(r-1)N_f$ directions provide the theta functions
multiplying $h_{r,c_1}(z,\tau; J_{0,1},\IB_0)$. 

The parameter $w$ does not appear in the elliptic genus, therefore one
should specialize to numerical invariants by taking the limit $w\to
-1$. For example for $\IF_0$ one finds for $J_{\varepsilon,1}$:
\begin{eqnarray}
\label{eq:htaueps}
&&h_{2,\beta \bC-\alpha \bff}(\tau;J_{\varepsilon,1})=0,\qquad \beta=1\mod 2 \non \\
&&h_{2,\beta \bC-\alpha
  \bff}(\tau;J_{\varepsilon,1})=-\frac{E_2(\tau)-1}{12\,\eta(\tau)^8},\qquad
(\alpha,\beta)=(1,0)\mod 2, \\
&&h_{2,\beta \bC-\alpha 
  \bff}(\tau;J_{\varepsilon,1})=-\frac{E_2(\tau)+2}{12\,\eta(\tau)^8},\qquad
(\alpha,\beta)=(0,0)\mod 2,\non
\end{eqnarray}
where $E_2(\tau)$ is the Eisenstein series of weight 2. For $J_{1,1}$, the coefficients can be expressed in terms of class
numbers $H(n)$ (which count binary quadratic forms
with given discriminant \cite{zagier:1975}):
\begin{eqnarray}
\label{eq:htau11}
h_{2,\bC - \bff}(z,\tau;J_{1,1})=\frac{\vartheta_2(\tau) \sum_{n\geq 0} H(-1+8n)\, q^{n}}{\eta(\tau)^8}.
\end{eqnarray}

An important aspect of the functions above is that they do not quite 
transform as a modular form of weight $-\chi(\IB_n)/2=-2-n/2$ as the
partition function for $r=1$. This is easily seen from
Eqs. (\ref{eq:htaueps}) and (\ref{eq:htau11}), since $E_2(\tau)$ and
the class number generating function\footnote{The notation here means to sum over $n$ from $0$ to $\infty$ where $n$ is either $0$ or $3$ modulo $4$.}
$\mathfrak{h}(\tau)=\sum_{n=0 \atop n=0,3 \mod 4}^\infty H(n)q^n$ transform only as a modular form
after addition of a suitable non-holomorphic term:
\be
\label{eq:mockexamples}
\hat E_2(\tau)=E_2(\tau)-\frac{3}{\pi \im(\tau)}, \qquad
\mathfrak{\hat h}(\tau)=\mathfrak{h}(\tau)+\frac{(1+i)}{8\pi}\int_{-\bar \tau}^{i\infty}\frac{\vartheta_3(u)}{(\tau+u)^{3/2}}du. 
\ee
Using the techniques of \cite{Zwegers:2000}, the required
non-holomorphic terms can be derived for general values of $J$ and
also $w\neq -1$.

Our proposal states that the partition functions above appear as the
elliptic genus of a (0,4) sigma model. For $r\geq 2$, however, new issues
appear which make the proposal more involved. An important issue
is how the dependence of  $h_{r,c_1}(\tau;J)$ on $J$ for $r>1$ is
realized on the two-dimensional side. The partition functions for weak
coupling or equivalently $J=J_{\varepsilon,1}$ do not have a form which is familiar from
conformal field theory. On the other hand, Appell-Lerch functions
which appear for $J=-K_{\IB_n}$ do appear as characters in conformal
field theory \cite{Eguchi:1988af, Kac:2000, Semikhatov:2003uc}, and
interestingly also as partition functions of CFT's 
with a non-compact field space \cite{Troost:2010ud}. Non-holomorphic
terms as the integral in (\ref{eq:mockexamples}) are argued to be a
direct consequence of the (regularization of) non-compact target space. This is
in nice agreement with our findings, since the Atiyah-Hitchin moduli
space is also non-compact. The structure of the $\cN=4$ Yang-Mills partition function therefore
indicates that the elliptic genus of the (0,4) CFT corresponds to the polarization
$J=-K_{\IB_n}$, and that the two-dimensional field theory dual to
Yang-Mills theory for other values of $J$ is not conformal. Around
(\ref{eq:basisEn}) is explained that this point $J=-K_{\IB_n}$ is also
very special in the five-dimensional theory, where it corresponds to
$E_{n+1}$ RG fixed point at infinite coupling. Note that the jumps in the quantum spectrum of $\cN=4$ Yang-Mills
correspond in the 2d theory to changes of the spectrum along its
RG-flow. The Yang-Mills partition function provides an interesting
tool for studying such flows.

The above point of view is also consistent with the attractor flow of the
moduli in five-dimensional supergravity solutions sourced by
M5-branes. In the near horizon AdS$_3$ region, the moduli are fixed at
their attractor point.  To see the attractor flow one needs to move
out of the near-horizon region. From the point of view of the dual
conformal field theory, perturbing the moduli away from the attractor
point corresponds to an irrelevant perturbation of the conformal field
theory \cite{deBoer:2008ss}. In our local Calabi-Yau manifold, the
attractor point determines $J$ to be proportional to
$-K_{\IB_n}$. (Recall that for $\cN=4$ Yang-Mills only the direction
of $J$ in the K\"ahler moduli space is relevant.) Thus
the supergravity viewpoint suggests that perturbing $J$ away from
$-K_{\IB_n}$ correspond to perturbing away from the IR fixed point.

In the above, we have concentrated on $r\leq 2$. For $r>2$ one
can also compute explicitly the (holomorphic part of 
the) partition functions of $\cN=4$ Yang-Mills. The indefinite theta
 functions are rather involved \cite{Manschot:2010nc,
   Manschot:2011dj}. It would be very interesting to relate these
 functions to those of a conformal field theory.

\subsection{Relation with $\cN=2$ Yang-Mills theory in four dimensions}
\label{sec:4dimensions}
Compactification of the five-dimensional theory on a circle with radius $R_2\to 0$
gives in four dimensions $\cN=2$, $SU(2)$ gauge theory with $N_f$ hypermultiplets.  
The magnetic string of the five-dimensional theory
corresponds to a magnetic monopole or dyon preserving half of the $\cN=2$
supersymmetry of the four dimensional theory. The spectrum of
these BPS-monopoles and dyons is fully known. We point out in this section that these
spectra are consistent with the partition functions of $\cN=4$ Yang-Mills
computed in Subsection \ref{subsec:N4YM}, in agreement with the ``no
walls'' conjecture of Ref. \cite{Chuang:2013wt}. 

The monopole and dyon spectrum of $\cN=2$ gauge theory with $N_f$
flavors can in principle be determined by computing the Dirac index of the monopole
moduli space twisted by the connection coming from the flavor
fermions \cite{Sethi:1995zm,Cederwall:1995bc,Gauntlett:1995fu}, and is
consistent with later analysis \cite{Ferrari:1996sv, Bilal:1996sk,
  Alim:2011kw}. We list here the spectra for $N_f\leq 3$. The four-dimensional magnetic and electric
charge is denoted by $(n_m,n_{4,e})$. The W-boson has for any $N_f$ charge $(0,2)$, and is a singlet of the $SO(2N_f)$ flavor
group. For $N_f=0$, the spectrum consists furthermore of the
infinite set of dyons with charge $(1,2n),\,n\in \mathbb{Z}$ \cite{Seiberg:1994rs}. The BPS-index $\Omega$ (at
weak coupling) of the W-boson is equal to $-2$, and that of the
monopole and dyons is equal to 1 \cite{Gaiotto:2008cd}. 
For $N_f=1$, the spectrum of the quarks and dyons is given by
(see e.g. \cite{Alim:2011kw}):
\begin{center}
\begin{tabular}{l|r}
 Particle $(n_m,n_{4,e})$ & $SO(2)$ charge \\
\hline
Quarks $(0, 1)$ & $\pm 1$  \\
 Dyons $(1, 2n)$  & $\frac{1}{2}$  \\
$\qquad \quad (1, 2n+1)$ &  $-\frac{1}{2}$
\end{tabular}
\end{center}
for $N_f=2$:
\begin{center}
\begin{tabular}{l|c}
Particle $(n_m,n_{4,e})$ & Rep. $SO(4)\cong SU(2)\otimes SU(2)$ \\
\hline
Quarks $(0, 1)$ & $\bf (2,2)$ \\
Dyons $(1, 2n)$ & $\bf (2,1)$ \\
$\qquad \quad (1, 2n+1)$ & $\bf (1,2)$
\end{tabular}
\end{center}
and for $N_f=3$:
\begin{center}
\begin{tabular}{l|c}
Particle $(n_m,n_{4,e})$ & Rep. $SO(6)\cong SU(4)$ \\
\hline
Quarks $(0, 1)$ & $\bf 6$ \\
Dyons $(1, 2n)$ & $\bf 4$ \\
 $\qquad \quad  (1, 2n+1)$ & $\bf \bar 4$ \\
$\qquad \quad (2,2n+1)$ & $\bf 1$ 
\end{tabular}
\end{center}

The compactification $R_2\to 0$ corresponds to the well-known limit of
M-theory giving Type IIA string theory. The four-dimensional $\cN=2$,
$SU(2)$ gauge theory is engineered by taking a double scaling 
limit in the K\"ahler moduli space of the non-compact Calabi-Yau
\cite{Katz:1996fh}. One can therefore arrive at this
spectrum by continuation of semi-stable sheaves, which are the mathematical
description of BPS-states in the large volume limit, to the field
theory regime of the K\"ahler moduli space. The magnetic charge
is given by the rank of the sheaf: $n_m=r$ as in five dimensions. However in order
to reproduce the spectra in the tables, one finds that the
4-dimensional electric charge $n_{4,e}$ is shifted with respect to the
5-dimensional charge $n_e$:  
\be
n_{4,e}=n_e- n_m \,N_{f}/2.
\ee
As a result the relation between the first Chern class, $r$ and the
electric charge is given by $\bfk=c_1-rK_{B_{N_f}}/2=\half (n_{4,e}+r N_f/2) \bff-\half
\sum_{i}n_{f,i}\,\bd_{i}$. With this identification, the electric
charge $n_{4,e}$ is an integer as in the literature. One can easily verify that the DSZ
symplectic innerproduct is independent of this shift: $\left< (n_m,n_{4,e}),(n_m',n_{4,e}')
\right>=n_m n_{4,e}'-n_{4,e}n_m'=\left< (n_m, n_e),(n_m', n_e')
\right>$. 

Note that there are no charges in the field
theory corresponding to D0-brane (second Chern character)  or
D2-brane (first Chern character)  supported on $\bC$, since these objects become very massive
in the field theory limit and leave the spectrum.
Therefore, at most the lowest term in the $q$-expansion of the $\cN=4$ Yang-Mills partition functions
correspond to monopoles and dyons in the field theory. To
determine which of the lowest term indeed represent BPS-states of the
field theory, one has to verify that their mass is at the field theory
scale and not of string scale, and that furthermore no walls of marginal stability are crossed by the K\"ahler moduli in
between the large volume limit and the field theory limit. This
analysis is carried out for pure $SU(2)$ gauge theory in
\cite{Chuang:2013wt} and conjectured to be generically valid for gauge
theories. In the following, we will confirm the
conjecture for $SU(2)$ gauge theory with $N_f\leq 3$ by matching the BPS
spectra with the BPS invariants of semi-stable sheaves.

The correlation between electric and flavor charges
\cite{Seiberg:1994aj, Gauntlett:1995fu} can be seen from
this perspective as a natural consequence of the gluing vectors of the
lattice $A\oplus D$. The W-boson lies in the conjugacy class
of $\bg_0$, and the quarks in the conjugacy class of
$\bg_2$. The two other conjugacy classes $\bg_{1,3}$ are not relevant for
$r=0$ due to the large size of the curve $\bC$. Since 
the charge vector $\bfk$ differs from the first Chern class $c_1$ by
$rK_S/2$, the electric and flavor charges of monopoles with $n_m=r=1$
take values in the classes $\bg_{1,3}$. The dimension of the $SO(2N_f)$
representation of the monopoles and dyons are provided by the first
coefficients of the theta functions (\ref{eq:Dnthetas}). These indeed agree with the
dimensions obtained in the references listed in the above tables. 
To further verify that the dyons lie in hypermultiplets whose
BPS index is 1, we need to consider the functions
$h_{1,c_1}(z,\tau;\IB_{N_f})$. The generating functions for $r=1$ and any
choice of $c_1$ have the expansion:  
\be
h_{1,c_1}(z,\tau;\IB_{N_f})=\frac{q^{-\frac{4+N_f}{24}}}{w-w^{-1}}\left(1+(w^{-2}+2+N_f+w^2)\,q+\dots\right),
\ee
which indeed confirms that the dyons lie in hypermultiplets.  

With the results of Section \ref{sec:rgeq2}, we can also address the
monopole with magnetic charge 2, which is part of the BPS-spectrum of SYM
with $N_f=3$ \cite{Seiberg:1994aj,Sethi:1995zm,
  Gauntlett:1995fu}. Since $r=2$ and the electric charge $n_{4,e}$ is
odd, $c_1=\half (n_{4,e}+3)\bff+K_{B_{N_f}}$
lies in the conjugacy class $\bg_0$. The expansion of
$\Theta_{D_3, \bg_0}$ starts with 1 (\ref{eq:Dnthetas}), which is in agreement
with this dyon being a singlet of $SO(6)$. In order to further verify
that a semi-stable sheaf exists with BPS invariant equal to $1$, we expand
$h_{2,c_1}(z,\tau;\IB_3,J_{\varepsilon,1})$ (\ref{eq:h23b}) for
$c_1=\half (n_{4,e}+3) \bff+K_{\IB_3}$ and $n_{4,e}$ odd: 
\be
h_{2,c_1}(z,\tau;\IB_3,J_{\varepsilon,1})=\frac{q^\frac{1}{6}}{w-w^{-1}}\left(1+(w^{-4}+6w^{-2} +16+6w^2+w^4)\,q+\dots\right).
\ee
Indeed, this expansion starts with 1, and we have thus confirmed the existence of the $SO(6)$ singlet dyon
with magnetic charge 2 from sheaf counting in the large volume limit. 

\section{Discussion and outlook}
We have proposed a correspondence between $(0,4)$ sigma models with target
space the moduli spaces of $r$ static monopoles in $SU(2)$
four-dimensional gauge theory, and $\cN=4$ $U(r)$ Yang-Mills
theory on the four-manifolds known as del Pezzo surfaces. This correspondence can be understood
from the point of view of geometric engineering of five-dimensional
gauge theory by M-theory compactified on a non-compact
Calabi-Yau manifold. The del Pezzo surfaces form the compact part of
these Calabi-Yau manifolds. For $r=1$, we have proven the
correspondence, while for higher rank $r\geq 2$, much work remains to
be done. Clearly, the computation of the 2d elliptic genus for $r>2$
similarly to the 4d SYM computation is desirable. Another important missing point is the relation of
the sigma model of the monopole moduli space considered in this paper, 
with the field theory obtained from the reduction of the degrees of
freedom to two dimensions of multiple M5-branes wrapping $\IB_0$ along
the lines of \cite{Maldacena:1997de,Gaiotto:2006wm,Minasian:1999qn} for the case of
a single fivebrane and compact Calabi-Yau threefolds. One might expect
that for $r\geq 2$, the sigma model considered in this paper can be derived as a
Coulomb branch of a more complicated field theory, analogously to the
Coulomb phase of the quiver quantum mechanics describing
BPS bound states \cite{Denef:2002ru}. 

We have left unexplored various aspects of the 5d/2d/4d correspondence. For
example, to consider the more general class of four-manifolds which
are $\IP^1$ fibrations over a genus $g$ Riemann surface instead of
over a $\IP^1$. This would lead in the engineered $SU(2)$ gauge theory to $g$
hypermultiplets in the adjoint representation \cite{Witten:1996qb},
whereas in our discussion we restricted to blow-ups of $\IP^1\times
\IP^1$, leading to hypermultiplets in the fundamental representation. Also, we have not included
possible mass terms for the hypers in our analysis, apart from some rather simple observations made in Sections 2 and 3. 
Including them in the elliptic genus is an interesting extension. Higher rank gauge
groups $SU(N>2)$ in five-dimensional gauge theory can also be considered. They could lead to new versions of the 5d/2d/4d correspondence.

Another interesting direction for future research is the study of instanton effects of the
three-dimensional theory obtained from the five-dimensional theory
compactified on $T^2$ \cite{arXiv:1107.2847}. These corrections contribute to the 
hypermultiplet moduli space metric on the Coulomb branch of the three-dimensional effective action. 
The hypermultiplet moduli space was shown to be equal to the moduli space of doubly periodic monopoles in \cite{Cherkis:2012qs}, but the metric remains difficult to be computed. In \cite{arXiv:1107.2847}, it was argued that for $N_f=0$, the elliptic genus determines the instanton induced four-fermi correlator in the three-dimensional gauge theory, and hence the hypermultiplet moduli space metric. 
Furthermore, these corrections are beautifully captured using integrals over twistor space
\cite{Gaiotto:2008cd, Alexandrov:2008gh,Crichigno:2012vd}, and are
expected to be invariant under $SL_2(\mathbb{Z})$ transformations of
the $T^2$. Progress on these aspects for one-instanton
corrections is recently made in \cite{Alexandrov:2012au}. An
intriguing interplay is expected for higher
instanton corrections between the period integrals as in
Eq. (\ref{eq:mockexamples}) and twistor integrals.  
   
Finally, it would be worth investigating five-dimensional gauge theories, possibly at the superconformal point, on different manifolds, such as $S^3\times T^2$, to see if a 5d/2d/4d correspondence still holds. If so, connections could be made to the study of partition functions of five-dimensional gauge theories using localization techniques, along the lines of \cite{Kallen:2012cs,Hosomichi:2012ek,Kim:2012gu,Jafferis:2012iv,Lockhart:2012vp}. We leave this for future investigation.

\subsection*{Acknowledgments}
We thank C. Cordova, E. Diaconescu, A. Klemm, G. Moore, V. Pestun, B. Pioline and C. Vafa for useful
discussions. J.M. is grateful for hospitality of the CEA Saclay, the Hausdorff Research Institute,
Harvard University, the Isaac Newton Institute and Utrecht University
during various stages of this work. B.H. and S.V. would like to thank the 2012 Simons workshop in Mathematics and Physics and the Simons Center for Geometry and Physics for hospitality during an important stage of this work. B.H. would further like to thank Utrecht University and University of Bonn for hospitality during various stages of this work. The work of B.H. is supported by DFG fellowship HA 6096/1-1. Further financial support is given by the Netherlands Organization for Scientific Research (NWO) under the VICI grant 680-47-603.


 \end{document}